\documentclass[journal]{IEEEtran}
\usepackage{cite}
\usepackage{amsmath}
\usepackage{amssymb}  % assumes amsmath package installed
\usepackage{accents}
\usepackage{mathtools}
\DeclareMathAlphabet{\mathpzc}{OT1}{pzc}{m}{it}
\newcommand{\goodchi}{\protect\raisebox{2pt}{$\chi$}} 	
\usepackage{algorithmic}
\usepackage{algorithm}
\usepackage{array}
\usepackage{url}
\usepackage[bookmarks=false]{hyperref}
\usepackage{subfigure}
\usepackage{xcolor}
\newcommand{\pt}{\color{black}}
\newcommand{\TR}[1]{{\color{black} #1}}

\newcommand{\REV}[1]{{\textcolor{black}{#1}}}

\title{Gaussian-Process-based Adaptive Tracking Control with Dynamic Active Learning for Autonomous Ground Vehicles}

\author{Kristóf~Floch, Tamás~Péni, and~Roland~Tóth\thanks{The authors are with the Systems and Control Laboratory, \mbox{HUN-REN} Inst. for Computer Science and Control, Budapest, Hungary. R. Tóth is also affiliated with the Control Systems Group, Eindhoven University of Technology, Eindhoven, The Netherlands and the Vehicle Research Center, Sz\'{e}chenyi Istv\'{a}n University, Gy\H{o}r, Hungary. K. Floch is also affiliated with ETH Zürich, Zürich, Switzerland. (e-mails: 
{\tt\scriptsize kfloch@ethz.ch,peni@sztaki.hun-ren.hu,r.toth@tue.nl})\\ This work was supported by the European Union within the Framework of the National Laboratory for Autonomous Systems under Grant RRF-2.3.1-21- 2022-00002.}}

\begin{document}

\maketitle

\begin{abstract}
This article proposes an active-learning-based adaptive trajectory tracking control method for autonomous ground vehicles to compensate for modeling errors and unmodeled dynamics. The nominal vehicle model is decoupled into lateral and longitudinal subsystems, which are augmented with online Gaussian Processes (GPs) using measurement data. The estimated mean functions of the GPs are used to construct a feedback \TR{compensator}, which, together with an LPV state feedback controller designed for the nominal system, gives the adaptive control structure. To assist exploration of the dynamics, the paper proposes a new, dynamic active learning method to collect the most informative samples to accelerate the training process. To analyze the performance of the \TR{overall learning tool-chain provided controller}, a novel iterative, counterexample-based algorithm is proposed for calculating the induced $\mathcal{L}_2$ gain between the reference trajectory and the tracking error. \TR{The analysis can be executed for a set of possible realizations of the to-be-controlled system, giving robust performance certificate of the learning method under variation of the vehicle dynamics.} The efficiency of the proposed control approach is shown on a high-fidelity physics simulator and in real experiments using a 1/10 scale F1TENTH electric car. 

\end{abstract}
\vspace{-5mm}

%%%%%%%% PAPER %%%%%%%%%
\section{Introduction}
\IEEEPARstart{N}{owadays}, autonomous mobile robots, such as small-scale wheeled ground vehicles are becoming wide-spread in various industrial applications, hence, further improvement of autonomous maneuvering capabilities that can exploit the motion dynamics of these robots is a subject of scientific research. An essential prerequisite for reaching general utilization of these vehicles is the development of simple but effective control algorithms that can cope with unknown variations in the motion dynamics and can ensure high-performance maneuvering with formal guarantees of stability and performance.

The state-of-the-art trajectory tracking algorithms for ground vehicles are usually studied in the context of autonomous racing \cite{Betz22_autonomous_racing}, utilizing both model-free \cite{balaji2019deepracer} and model-based \REV{\cite{Liniger_Domahidi_Morari_2015, Becker23_MAP, Rosolia17_LMPC}} approaches. While model-free methods have gained significant attention, in the context of autonomous vehicles, model-based solutions are still often favored due to safety concerns. The performance of model-based approaches is dependent on the accuracy of the underlying model, therefore, precise identification of the motion dynamics is required. % for their deployment.
For this purpose, often first-principles-based models are used, but even after estimating the parameters of such models, unknown and hard-to-model dynamic effects that are specific to each vehicle often have significant effects on the behavior of the system. As a result, model uncertainties are inevitable and can lead to performance degradation in precise maneuvering in practice. To efficiently mitigate the effect of modeling uncertainties, adaptive algorithms have been extensively studied, which either estimate the parameters of the system dynamics, \cite{Vaskov22_friction_adaptive}, or directly adapt the parameters of a control policy \cite{Kebbati21_AMPC}. Furthermore, model augmentation, where a nominal vehicle model, typically based on first principles, is complemented with a learning component to capture the residual model mismatch, has been investigated. These methods usually utilize machine learning in terms of \emph{artificial neural networks} (ANN) \cite{Xuewu18_ann}, or \emph{Gaussian processes} (GP) \cite{Kabzan19_GPMPC} to capture the unknown model dynamics. %{as these techniques are also capable of handling variation in the model structure.
GP augmentation has proven to be beneficial for a wide range of mobile robotic applications such as cars \cite{Hewing18_MPCGP} quadcopters \cite{liu22_learning} and robotic arms \cite{Carron19_gpmpcarm}, due to its high approximation capability and the uncertainty characterization of the estimates. 

For GP-augmented models, \emph{nonlinear model predictive control} (NMPC) algorithms are usually  preferred \cite{Kabzan19_GPMPC, Hewing18_MPCGP, liu22_learning, Carron19_gpmpcarm}. However, due to the online optimization, real-time NMPCs  require substantial computational resources compared to classical feedback methods, \REV{therefore, the latter is still favored in applications}, where processing power is limited \cite{Becker23_MAP}. Therefore, this article proposes an effective GP-based \REV{feedback} compensation method with low computational complexity to adapt to modeling uncertainties.

Collecting measurement data for the training of the GPs in real-world scenarios is crucial for the effectiveness of the learning methods and it can be a time and energy-consuming task. Often, this step is overlooked in the literature. Furthermore, constraints of the specific environments (e.g.\ limited test areas, obstacles) and the motion dynamics of the vehicle make the data acquisition challenging. Active learning methods have been a popular choice for the identification of unknown dynamics \cite{kiss2024_spacefilling} and to refine GP models \cite{Krause08_GP, Gango19_GP_tuning} using their posterior distribution. However, existing methods assume that input-output training data can be easily collected, i.e., the function to be approximated can be evaluated at any value of its arguments. In our case, the inputs of the GP are the {\pt states of the motion dynamics}. Therefore, to collect training data, the dynamic model has to be driven to the required state configuration. To address this issue, we propose a novel dynamic active learning approach that takes into account the system dynamics and systematically designs experiments for data collection.

%Once the training dataset is obtained, GPs are usually trained offline, with gradient-based methods \cite{Rasmussen05_GP}. As the computational complexity training the GPs is cubic in terms of the training set, GPs are generally not capable of handling large training datasets. To counter this issue several scalable GP methods have been developed with the goal of reducing the computational cost while preserving accuracy. The most advanced techniques are successfully utilized in robotic applications are the \emph{fully independent training conditional} (FITC) \cite{Snelson2005_SparseGP_FITC} and the \emph{variational free energy} (VFE) \cite{Titsias09_sparseGP}. To have an adaptive control algorithm that can tune itself during operation, the online update of the GP models is necessary. However, most previously outlined solutions do not utilize online updates to avoid data bloating. To counter this issue various online update schemes have been proposed such as the \emph{recursive least squares} (RLS) \cite{liu22_learning, Liu24_Attitude} or the distance-based dictionary update method \cite{Kabzan19_GPMPC}.

\REV{Providing performance guarantees is a crucial requirement in the study of autonomous systems. In practice, feedback algorithms for autonomous vehicles often rely on decoupled or simplified dynamic models \cite{Gagliardi24_LPVl2, Gupta19_lateral_framework}, or are based solely on kinematic representations \cite{Paden16_survey}. As a result, these approaches cannot ensure stability of the full nonlinear closed-loop system by design. Guaranteeing stability for nonlinear systems is inherently challenging, particularly when learning-based components, such as GPs or ANNs, are included in the control loop \cite{Morari21_ROA}.}

To summarize, the main expectations towards control algorithms for such vehicles are (a) adaptability to epistemic uncertainties, unknown dynamics and changing environmental conditions; (b) computationally efficient implementation; (c) guaranteed stability and performance. %in the operating range.
To address the presented challenges, our main contributions\footnote{A preliminary version of the proposed control architecture has been published in \cite{Floch24_ECC}. This article extends \cite{Floch24_ECC} by the online training of the GPs, the experiment design by active learning, extended performance guarantees, and real-world experimental evaluation.} are as follows:
\begin{enumerate}
    \item[C1] We propose a computationally efficient GP-based adaptive trajectory tracking control architecture for car-like platforms, capable of handling large model mismatch. %{The controller relies on a  dynamic vehicle model augmented with GP-based learning components to estimate model uncertainties.}
    \item[C2] We propose a recursive online sparse GP update technique, capable of jointly updating the hyperparameters and the inducing points of the GP.
    \item[C3] We propose a novel active learning method for dynamic systems that utilizes the posterior distribution of the GPs to synthesize trajectories along which the learning components can be refined in a dynamic setting.
    \item[C4] We provide a robust analysis \TR{approach} in terms of the worst-case achievable $\mathcal{L}_2$ gain \TR{of the overall learning toolchain provided GP compensator under variations of the to-be-regulated vehicle dynamics}.
    \item[C5] We demonstrate the performance of the proposed adaptive methods in a high-fidelity simulator and in real experiments using the F1TENTH platform \cite{OKelly20_f1tenth}.
\end{enumerate}

The remainder of the paper is organized as follows. Sec.~\ref{sec:GP} provides an overview of GP regression and the proposed online recursive update method. It is followed by the problem formulation and the discussion of the baseline model in Sec.~\ref{sec:model_and_problem}. Sec.~\ref{sec:adaptive_control} presents the proposed learning-based trajectory tracking algorithm, followed by dynamic active learning in Sec.~\ref{sec:BO-planning}, Finally, the simulation study is presented in Sec.~\ref{sec:simulations}, while the experimental results are described in Sec.~\ref{sec:implementation}.

\vspace{-3mm}
\section{Gaussian Process Regression}
\label{sec:GP}
\emph{Gaussian processes} (GPs) have been a popular choice for model augmentation, see \cite{Kabzan19_GPMPC,Hewing18_MPCGP}, as they are efficient function estimators with direct characterization of the uncertainty of the estimate \cite{Rasmussen05_GP} and a manageable computational load compared to Bayesian neural networks. Compared to most of the other learning approaches such as ANNs, the uncertainty of the approximation can be employed for designing controllers that guarantee robust performance of closed-loop despite the uncertainty of the model of the underlying behavior of the plant \cite{Antal23_backflip}. Next, we give a brief overview of GP-based modeling and techniques that enable real-time implementation. 

\vspace{-3mm}
\subsection{Gaussian Processes}
To estimate a scalar unknown function $f: \mathbb{R}^{n_\mathrm{x}}\rightarrow \mathbb{R}$ with a GP, let $X^N=[\REV{\mathpzc{x}}_1^\top\;\cdots\;\REV{\mathpzc{x}}_N^\top]\in\mathbb{R}^{N\times n_\mathrm{x}}$ denote the collection of $N$ number of input values and let $Y^N=[\REV{\mathpzc{y}}_1\;\cdots\;\REV{\mathpzc{y}}_N]^\top\in\mathbb{R}^{N}$ be the corresponding noisy output measurements, forming the data set $\mathcal{D}^N=\{X^N,Y^N\}$, generated by the following model:
\begin{equation}
\label{eqn:gp_model}
    \REV{\mathpzc{y}}_i=f(\REV{\mathpzc{x}}_i)+\epsilon_i, \quad i\in\mathbb{I}_1^N,
\end{equation}
where $\epsilon_i \sim \mathcal{N}(0,\sigma_\epsilon^2)$ is an i.i.d. Gaussian noise and $\mathbb{I}_{\tau_1}^{\tau_2}=\{i\in\mathbb{Z}\;|\;\tau_1\leq i \leq \tau_2 \}$.
The core idea of GP-based estimation of $f$ is to consider that candidate estimates $g$ belong to a GP, seen as a prior distribution. Then, using $\mathcal{D}_N$ and this prior, a predictive GP distribution of $g$ is computed that provides an estimate of $f$ in terms of its mean and describes the uncertainty of this estimate by its variance.

 In terms of definition, a scalar-valued \emph{Gaussian Process} $\mathcal{GP}$ assigns to every point $\REV{\mathpzc{x}}\in\mathbb{R}^{n_\mathrm{x}}$ a random variable $\mathcal{GP}(\REV{\mathpzc{x}})$, such that, for any
finite set $\{\REV{\mathpzc{x}}_i\}_{i=1}^{N}\subset\mathbb{R}^{n_\mathrm{x}}$, the joint probability distribution of $\mathcal{GP}(\REV{\mathpzc{x}}_1),\cdots,\mathcal{GP}(\REV{\mathpzc{x}}_N)$ is Gaussian. Due to this property, $g\sim \mathcal{GP}(m,\kappa)$ is fully determined by its mean $m$ and covariance $\kappa$ expressed as
\begin{subequations}
    \begin{align}
        m(\REV{\mathpzc{x}})&=\mathbb{E}[g(\REV{\mathpzc{x}})],\\
        \kappa(\REV{\mathpzc{x}},\tilde{\REV{\mathpzc{x}}})&=\mathbb{E}[(g(\REV{\mathpzc{x}})-m(\REV{\mathpzc{x}}))(g(\tilde{\REV{\mathpzc{x}}})-m(\tilde{\REV{\mathpzc{x}}}))],
    \end{align}
\end{subequations}
where $\REV{\mathpzc{x}},\tilde{\REV{\mathpzc{x}}}\in\mathbb{R}^{n_\mathrm{x}}$ and $\mathbb{E}$ is the expectation.
This distribution describes our prior belief of the function space in which the estimate of the   %our prior belief of
the unknown function $f$ is searched for. %For the estimation, 
In the remainder of the paper, we assume w.l.o.g. that the prior mean is zero ($m(\REV{\mathpzc{x}})=0$) and covariance of the distribution is characterized with a \emph{squared exponential} (SE) kernel, a common choice for estimation of smooth functions:
\begin{equation}
    \kappa_\mathrm{SE}(\REV{\mathpzc{x}},\tilde{\REV{\mathpzc{x}}})=\sigma_\mathrm{f}^2\exp\left(-\frac{1}{2}(\REV{\mathpzc{x}}-\tilde{\REV{\mathpzc{x}}})^\top \Lambda^{-1}(\REV{\mathpzc{x}}-\tilde{\REV{\mathpzc{x}}})\right),
\end{equation}
where $\sigma_\mathrm{f}\in\mathbb{R}$ is a scaling factor and $\Lambda\in\mathbb{R}^{n_\mathrm{x}\times n_\mathrm{x}}$ is a positive definite and symmetric matrix that determines the smoothness of the candidate functions. The parameters of the considered kernel function, i.e.,  $\sigma_\mathrm{f}$, $\Lambda$ and $\sigma_\epsilon$ in \eqref{eqn:gp_model} are the so-called hyperparameters of the prior distribution and they are collected in the vector $\theta\in\mathbb{R}^{n_\theta}$. Note that other kernels can also be considered, as it is discussed in \cite{Rasmussen05_GP}.

%\cite{Rasmussen05_GP}.

%The joint probability distribution of $Y$ training output and an arbitrary test point $x_\star$ is given by
%\begin{equation}
%    \begin{bmatrix}
%        Y\\
%        g(x_\star)
%    \end{bmatrix} \sim \mathcal{N}\left(\begin{bmatrix}
%        0\\
%        0
%    \end{bmatrix},\begin{bmatrix}
%        K_{NN}+\sigma_\epsilon^2I & K_N(x_\star)\\
%        K_N^\top(x_\star) & \kappa(x_\star,x_\star)
%    \end{bmatrix}\right),
%
%\end{equation}
%where $[K_N(x_\star)]_i=\kappa(x_i,x_\star)$.

Based on $\mathcal{D}^N$ and the prior $g\sim \mathcal{GP}(m,\kappa)$,
\begin{equation}
    p(Y^N|X^N,\theta)=\mathcal{N}(0, K_{NN}+I\sigma_\epsilon^2)
\end{equation}
is the probability density function of the $Y^N$ outputs seen as random variables conditioned on $X^N$ and $\theta$, where $[K_{NN}]_{i,j}=\kappa(\REV{\mathpzc{x}}_i,\REV{\mathpzc{x}}_j),\; i,j\in\mathbb{I}_1^N$. Simple derivation leads to that the predictive distribution for $g(\REV{\mathpzc{x}}_\star)$ at test point $\REV{\mathpzc{x}}_\star$ is the posterior
%The resulting posterior distribution at an arbitrary test point $x_\star$ is also Gaussian with
$p(g(\REV{\mathpzc{x}}_\star) |\mathcal{D}^N,\REV{\mathpzc{x}}_\star)=\mathcal{N}(\mu(\REV{\mathpzc{x}}_\star),\Sigma(\REV{\mathpzc{x}}_\star))$, with
\begin{subequations}
\label{eqn:gp_mean_var}
\begin{align}
    \mu(\REV{\mathpzc{x}}_\star)&=K_N^\top(\REV{\mathpzc{x}}_\star)(K_{NN}+\sigma_\epsilon^2I)^{-1}Y^N\label{eqn:gp_mean},\\
    \Sigma(\REV{\mathpzc{x}}_\star)&=\kappa(\REV{\mathpzc{x}}_\star,\REV{\mathpzc{x}}_\star)-K_N^\top(\REV{\mathpzc{x}}_\star)(K_{NN}+\sigma_\epsilon^2I)^{-1}K_N(\REV{\mathpzc{x}}_\star),\label{eqn:gp_cov}
\end{align}
\end{subequations}
where $[K_N(\REV{\mathpzc{x}}_\star)]_i=\kappa(\REV{\mathpzc{x}}_i,\REV{\mathpzc{x}}_\star),\;i\in\mathbb{I}_1^N$. Eq. \eqref{eqn:gp_mean} characterizes the mean as the approximation of the unknown function $f$ and \eqref{eqn:gp_cov} is the variance, which gives the uncertainty %qualification 
of the approximation. 

Tuning the GP estimate, i.e., shaping the resulting posterior distribution, can be achieved by adjusting the previously introduced hyperparameters. An efficient approach for this follows via maximizing the marginal likelihood w.r.t.\ the hyperparameters, i.e.,
\begin{equation}
\theta^\star=\arg\max_{\theta}\{\log{p(Y^N|X^N,\theta)}\},  
\end{equation}
where
\begin{align}
\label{eqn:log_marginal_likelihood}
&\;\;\log{p(Y^N|X^N,\theta)}=-\frac{N}{2}\log(2\pi)\\&-\frac{1}{2}{Y^N}^\top(K_{NN}+\sigma_\epsilon^2I)^{-1}{Y^N}-\frac{1}{2}\log\det(K_{NN}+\sigma_\epsilon^2I).\nonumber
\end{align}
Commonly,\ \eqref{eqn:log_marginal_likelihood} is solved via standard gradient-based optimization such as conjugate gradient descent, providing an efficient way to optimize the hyperparameters.

As every training point is required for making predictions, the computational complexity of \eqref{eqn:gp_mean} is $\mathcal{O}(N)$ and of \eqref{eqn:gp_cov} is $\mathcal{O}(N^2)$, while the training is $\mathcal{O}(N^3)$ due to the evaluation \eqref{eqn:log_marginal_likelihood}. Therefore, in case of large training datasets, the real-time implementation of GP estimators is computationally expensive. 

\vspace{-3mm}
\subsection{Sparse Gaussian Processes}
\label{sec:SGP}
One way to tackle the computational limitation of GP regression is to use {\emph{sparse} GPs} (SGPs), where the number of data points used in the evaluation is limited to a fixed number. The goal of the SGP is to create a virtual dataset (often referred to as inducing points) defined as $\mathcal{D}^M=\{X^M,Y^M\}$, where $X^M=[\hat{\REV{\mathpzc{x}}}_{1}^\top \cdots \hat{\REV{\mathpzc{x}}}_M^\top]\in\mathbb{R}^{M\times n_x}$ and $Y^M=[\hat{\REV{\mathpzc{y}}}_1\cdots \hat{\REV{\mathpzc{y}}}_M]\in\mathbb{R}^{M}$ such that the Kullback-Leiber (KL) divergence between the posteriori distribution obtained from $\mathcal{D}^M$ and the original distribution based on $\mathcal{D}^N$ is minimal, while $M\ll N$.

%The core philosophy of the SGP is to select the set $\mathcal{D}^M$ and corresponding approximate Gaussian distribution $q\sim\mathcal{N}(\mathpzc{m},\mathpzc{K})$ in a manner that minimizes the Kullback-Leiber (KL) divergence between it and the distribution of the GP trained on the original dataset $\mathcal{D}^N$. 
To find the proper $\mathcal{D}^M$ set, we utilize the \emph{variational free energy} (VFE) approach \cite{Titsias09_sparseGP} in which a variational method is used that jointly selects the $X^M$ and the hyperparameters.

%The approximative posteriori distribution $\mathcal{N}(\mathpzc{m}, \mathpzc{K})$, i.e., the Nyström projection, can be calculated analytically as
%\begin{subequations}
%\label{eqn:sparse_approx_opt}
%    \begin{align}
%\mathpzc{m}&=\sigma^{-2}_\epsilon\mathpzc{K}K_{MM}^{-1}K_{MN}Y^N,\\        \mathpzc{K}&=K_{MM}%(K_{MM}+\sigma_\epsilon^{-2}K_{MN}K_{NM})^{-1}K_{MM}.
%    \end{align}
%\end{subequations}
%where $\mathpzc{m}$ is the mean of the approximation of the true posterior distribution and $ \mathpzc{K}$ is the associated covariance. Furthermore, $[K_{MN}]_{i,j}=\kappa({\hat{x}_i},x_{j}),\; i\in\mathbb{I}_1^{M},\; j\in\mathbb{I}_1^N$ is the covariance matrix between the pseudo inputs and all training inputs, $[K_{MM}]_{i,j},\; i,j=\mathbb{I}_1^M$ is the covariance of the pseudo inputs.

In the VFE approach, the objective can be expressed as 
\begin{multline}
\label{eqn:vfe}
%\mathcal{L}(\mathcal{D}^N,X^M, \theta)=\log(p(Y^N|X^N,X^M,\theta))\\-\frac{1}{2\sigma_\epsilon^2}\mathrm{trace}(K_{NN}-K_{NM}K_{MM}^{-1}K_{MN})%\\=&-\frac{1}{2}\big(N\log(2\pi)+{Y^N}^\top(K_{NM}K_{MM}^{-1}K_{MN}+\sigma_\epsilon I_N)^{-1}Y^N\\&+\log\det(K_{NM}K_{MM}^{-1}K_{MN}+\sigma_\epsilon I_N)\\&-\frac{1}{\sigma_\epsilon^2}\mathrm{trace}(K_{NN}-K_{NM}K_{MM}^{-1}K_{MN})\big).
\mathcal{L}(\mathcal{D}^N,X^M, \theta)= \frac{N}{2}\log(2\pi) + \frac{1}{2}Y^\top(W_{NN}+\sigma_\epsilon^{2}I_N)Y^N \\+ \frac{1}{2}\log\det(W_{NN} + \sigma_\epsilon^{2}I_N)+\frac{1}{2}\mathrm{tr}(K_{NN}-W_{NN})
\end{multline}
where $W_{NN}=K_{NM}K_{MM}^{-1}K_{MN}$ with $[K_{MN}]_{i,j}=\kappa({\hat{\REV{\mathpzc{x}}}_i},\REV{\mathpzc{x}}_{j}),\; i\in\mathbb{I}_1^{M},\; j\in\mathbb{I}_1^N$, which is the covariance matrix between the pseudo inputs and all training inputs, $[K_{MM}]_{i,j},\; i,j=\mathbb{I}_1^M$ is the covariance of the pseudo inputs. By maximizing \eqref{eqn:vfe}, the hyperparameters of the GP and the pseudo inputs can be jointly obtained, i.e.\
\begin{equation}
(\theta^\star, {X^M}^\star)=\arg\max_{\theta, X^M}\mathcal{L}(\mathcal{D}^N,X^M,\theta),   
\end{equation}
which is a nonlinear optimization that can be reliably solved. Furthermore, note that the pseudo outputs do not appear in the optimization, as they are fully represented by the approximative GP and the speudo-inputs.

After the training, the resulting posterior distribution at an arbitrary test point $\REV{\mathpzc{x}}_\star$ is $\mathcal{N}(\mu_\mathrm{p}(\REV{\mathpzc{x}}_\star),\Sigma_\mathrm{p}(\REV{\mathpzc{x}}_\star))$, with
\begin{subequations}
\label{eqns:sGP_pred}
    \begin{align}
        \mu_\mathrm{p}(\REV{\mathpzc{x}}_\star)&=K_{M}^\top(\REV{\mathpzc{x}}_\star)Q_{MM}K_{MN}\sigma^{-2}_\epsilon Y^N,\\
        \begin{split}
        \Sigma_\mathrm{p}(\REV{\mathpzc{x}}_\star)&=\kappa(\REV{\mathpzc{x}}_\star, \REV{\mathpzc{x}}_\star)-W_{MM}(\REV{\mathpzc{x}}^\star)+K_M^\top(\REV{\mathpzc{x}}^\star)K_{MM}^{-1}K_M(\REV{\mathpzc{x}}^\star),
        \end{split}
    \end{align}
\end{subequations}
where $[K_M(\REV{\mathpzc{x}}_\star)]_i=\kappa(\REV{\mathpzc{x}}_i,\REV{\mathpzc{x}}_\star),\;i\in\mathbb{I}_1^M$ and $W_{MM}(\REV{\mathpzc{x}}_\star)=K_M^\top(\REV{\mathpzc{x}}_\star)K_{MM}^{-1}K_M(\REV{\mathpzc{x}}_\star)$, which corresponds to a Nyström projection of the original GP to the pseudo inputs $X^M$. This formulation reduces the $\mathcal{O}(N^3)$ computational complexity of the evaluation of the mean to $\mathcal{O}(M)$ and the training to $\mathcal{O}(NM^2)$. In the following sections, SGPs will be utilized as the learning component of the proposed adaptive control approach.

\vspace{-4mm}
\subsection{Online Learning and Update for SGPs}
\label{sec:GP_update}
The VFE method is constructed for offline training, i.e., it requires the complete training set to be available. \REV{However, to provide adaptivity to changing environmental conditions, online training is necessary, which updates the model during operation using the latest measurement data.} For training GPs online, various algorithms such as the \emph{recursive least squares} GP (RLS-GP) \cite{liu22_learning} or the \emph{dynamic sparse} GP (DS-GP) \cite{Hewing18_MPCGP} are available. This article proposes a \emph{recursive gradient-based} (RGB) optimization scheme for SGPs utilizing the VFE method outlined in Sec.\  \ref{sec:SGP}. Compared to the previously introduced techniques that either update the approximate posterior distribution (RLS-GP) or the inducing points (DS-GP), this approach jointly updates both the hyperparameters and the inducing points of the GP based on the incoming new information, which, to the authors' knowledge, has not been applied before.

%Let the one incoming batch of input data at time step $k$ be denoted as $\mathcal{D}^Z_k=\{X^Z_k,Y^Z_k\}$, where $X^Z_k=[x_{1}^\top \cdots x_Z^\top]\in\mathbb{R}^{Z\times n_x}$, $Y^Z_k=[y_1\cdots y_Z]\in\mathbb{R}^{Z}$ and $Z$ is the batch size.% Then, one step of the recursive update based on $\mathcal{D}_k^Z$ can be formulated as
%\begin{equation}
%\label{eqn:rls_update_batch}
%    \begin{cases}
%     G_k= \lambda I + \Phi(X^Z_k) \mathpzc{K}_{k-1}\Phi^\top(X^Z_k),\\
%L_k=\mathpzc{K}_{k-1}\Phi(X^Z_k)G_k^{-1}
%\\
%\mathpzc{K}_k=\lambda^{-1}\left(\mathpzc{K}_{k-1}-L_kG_kL_k^\top\right)\\
%    r_k=Y^Z_k-\Phi(X^Z_k)\mathpzc{m}_{k-1}\\
%    \mathpzc{m}_k=\mathpzc{m}_{k-1}+L_kr_k
%    \end{cases}
%\end{equation}
%where $\Phi(X_k^Z)=K_M^\top(X_k^Z)K_{MM}^{-1}$, $0<\lambda\leq1$ is the forgetting factor of the RLS. The initial predictive mean $\mathpzc{m}_0$ can be obtained from an offline trained GP, based on an initial dataset, while for $\mathpzc{K}_0$ \cite{liu22_learning} suggests choosing $\mathpzc{K}_0=\beta^{-1}I_M$ with $0<\beta\leq1$. Smaller $\beta$ values indicate a lower confidence level for the trained GP, i.e.\ the function space to be learned is far away from the original training dataset, while higher $\beta$ assumes a smaller  difference. 

%\subsubsection{Gradient-based optimization (GRAD-GP)}
%\label{sec:gradsgp}
%In this section, we propose an optimization-based update method for SGPs, which can be used to efficiently update GP models online with incoming batches of new measurement data.
First, we assume that the SGP has been trained offline on the initial dataset $\mathcal{D}^N$, yielding the optimal values for $\theta$ hyperparameters and $X^M$ inducing points that correspond to $Y^M$ virtual outputs. As outlined before, the core idea of sparse GP estimation is that the virtual $\mathcal{D}^M$ dataset is obtained such that its distribution approximates the original $\mathcal{D}^N$ set with considerably fewer data points. Therefore, in future predictions, we can rely only on the $\mathcal{D}^M$ set, as it represents the information content of $\mathcal{D}^N$.

Let the one incoming batch of input-output data at time step $k$ be denoted as $\mathcal{D}^Z_k=\{X^Z_k,Y^Z_k\}$, where $X^Z_k=[\REV{\mathpzc{x}}_{1}^\top \cdots \REV{\mathpzc{x}}_Z^\top]\in\mathbb{R}^{Z\times n_\mathrm{x}}$, $Y^Z_k=[\REV{\mathpzc{y}}_1\cdots \REV{\mathpzc{y}}_Z]\in\mathbb{R}^{Z}$ and $Z$ is the batch size. Then, after receiving one batch of new training data at time $k$, we can combine the previously obtained pseudo dataset and the new batch to define a new training dataset as $\mathcal{\hat{D}}_k=\mathcal{D}^M_{k-1}\cup\mathcal{D}^Z_k$, where $\mathcal{D}_0^M=\mathcal{D}^M$. Note that with this formulation, $\mathcal{\hat{D}}_k$ both contains information from the pre-trained GP and innovation from the update batch. By substituting this combined dataset into the VFE cost \eqref{eqn:vfe}, we can perform $n_\alpha$ steps of gradient descent to collectively update the hyperparameters of the GP and the inducing points, i.e.\
\begin{equation}
\label{eqn:vfe_grad_desc}
    \begin{bmatrix}
        X^M_i \\
        \theta_i
    \end{bmatrix}=\begin{bmatrix}
        X^M_{i-1} \\
        \theta_{i-1}
    \end{bmatrix}-\alpha\frac{\partial \mathcal{L}}{\partial\begin{bmatrix}
        X^M_i \\
        \theta_i
    \end{bmatrix}}(\mathcal{\hat{D}}_k, X^M_i,\theta_i),
\end{equation}
where $\alpha$ is the learning rate and $i\in\mathbb{I}_1^{n_\alpha}$.
Note that the evaluation of \eqref{eqn:vfe_grad_desc} has the computational complexity of $\mathcal{O}((M+Z)M^2)$, therefore, gradient descent can be executed multiple times online in one update iteration. Furthermore, using the augmented training dataset $\mathcal{\hat{D}}_k$ is beneficial as it incorporates both the previously learned knowledge from $D^M_{k-1}$ and the new information contained in $\mathcal{D}_k^Z$. The RBG algorithm is outlined in Alg.\ \ref{alg:batch_update} and has three important parameters: size of the pseudo dataset $M$, update batch size $Z$ and $n_\alpha$ number of retrain iterations \REV{and $\alpha$ learning rate. The learning rate can be used to determine how much the GPs should favor new incoming data, i.e.\ how fast the GPs adapt to changing conditions. However, note that too fast adaptation can lead to forgetting, similarly to the RSL-GP.}

\begin{algorithm}[t]
\caption{Optimization-based batch update.}
\label{alg:batch_update}
\begin{algorithmic}[1]
\STATE \textbf{input:} $\mathcal{D}^M_{k-1}$ original pseudo dataset, $\theta_{k-1}$ hyperparameters, $\mathcal{D}^Z_k$ update batch, $\alpha$ learning rate, $n_\alpha$ number of gradient steps (retraining iterations)
\STATE $\mathcal{\hat{D}}_k \gets \mathcal{D}^M_{k-1} \cup \mathcal{D}^Z_k$
\STATE $\hat{X}^M_0 \gets X^M_k$
\STATE $\hat{\theta}_0 \gets \theta_k$
\FOR{$i \gets 1$ \TO $n_\alpha$}
    \STATE Calculate $\hat{X}^M_i$ and $\hat{\theta}_i$ using \eqref{eqn:vfe_grad_desc} with $\mathcal{\hat{D}}_k$, $\hat{X}^M_{i-1}$, $\hat{\theta}_{i-1}$
\ENDFOR
\STATE Update $\hat{\mathcal{D}}^M_k$ with $\hat{X}_{n_\alpha}^M$
\STATE $\theta_k \gets \hat{\theta}_{n_\alpha}$
\STATE \textbf{return} $\mathcal{D}^M_k$ and $\theta_k$
\end{algorithmic}
\vspace{-1mm}
\end{algorithm}
%The main benefit of the optimization-based batch update is that both the inducing points and the hyperparameters are updated in parallel at each iteration, unlike the RLS method, however, the computational overhead is also more significant.
\section{Vehicle Model \& Trajectory Tracking}
\label{sec:model_and_problem}

\subsection{Trajectory Tracking}
\label{sec:ref_traj}
The primary aim of this paper is to provide precise tracking of reference trajectories in the presence of modeling uncertainties. To outline the trajectory tracking problem, we first define the reference trajectories. The path is expressed as a two-dimensional spline curve $\psi(s^\mathrm{ref}(t))$, defined by the coordinate functions $(x(s^\mathrm{ref}(t)), y(s^\mathrm{ref}(t))$, where $s^\mathrm{ref}$ is a time domain signal defined as $s^\mathrm{ref}: \mathbb{R}\rightarrow[0,L]$. The arc length of the full path is denoted as $L$, hence $s^\mathrm{ref}(t)$ describes the desired vehicle position along the path at time $t$. Both $x(s^\mathrm{ref}(t))$, and $y(s^\mathrm{ref}(t))$ are \REV{continuous} in $s^\mathrm{ref}(t)$, furthermore $(x(0), y(0))$ and $(x(L), y(L))$ assign the endpoints of the curve. The speed reference $v^\mathrm{ref}(s^\mathrm{ref}(t))=v^\mathrm{ref}(t)$ along the trajectory is also given. These types of reference motion trajectories can be obtained by regular path planning algorithms.

%This section outlines the baseline first-principle vehicle model used in the remainder of the paper and then defines the trajectory tracking problem.
\vspace{-1mm}\subsection{Baseline Vehicle Model}
\label{sec:vehicle_model}
\begin{figure}%[thpb]
      \centering
      \includegraphics[width=.7\linewidth]{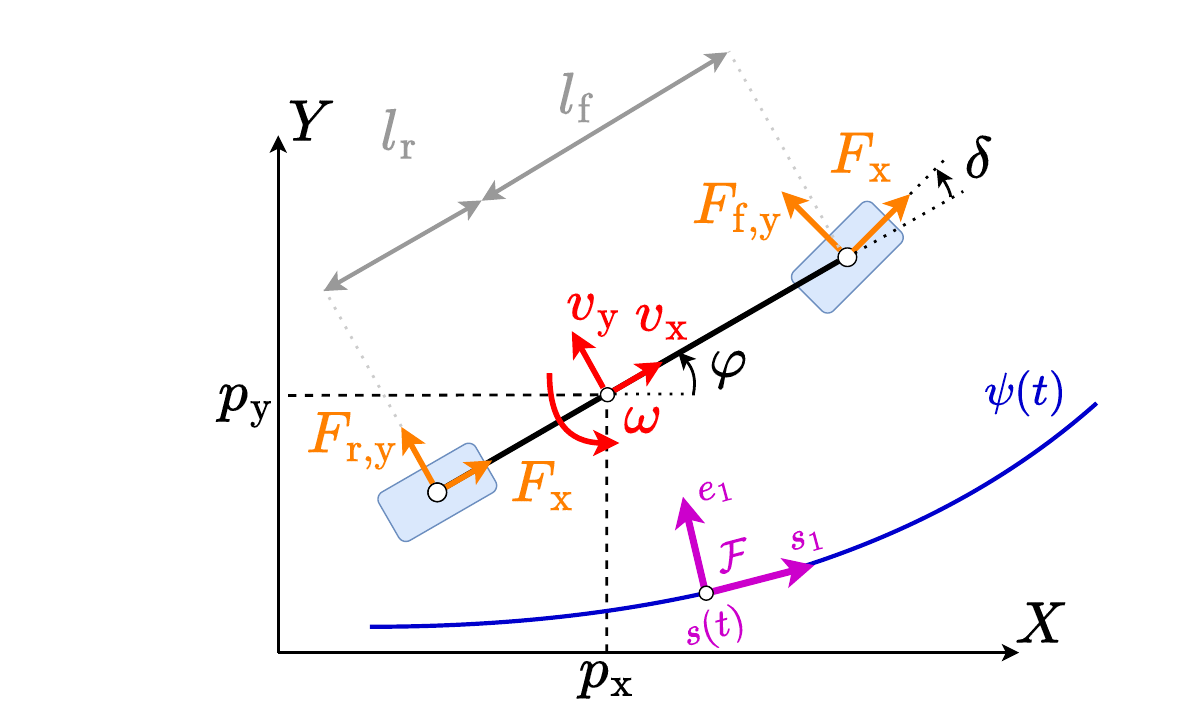}
      \caption{Single-track vehicle model and reference trajectory.}
      \label{fig:model}
      \vspace{-5mm}
   \end{figure}

The baseline vehicle model relies on a dynamic single-track representation,which has been commonly used for describing the behavior of small-scale car-like vehicles, see \cite{Floch22_f1tenth, Liniger_Domahidi_Morari_2015}. The modeling concept is depicted in Fig.\ \ref{fig:model} and the resulting model is described as
\begin{subequations}
\label{eqn:dyn_single_track_model}
\begin{align}
\REV{\dot{p}_\mathrm{x}}&=v_\mathrm{\REV{x}} \cos(\varphi)-v_\mathrm{\REV{y}} \sin(\varphi),\\
    \REV{\dot{p}_\mathrm{y}}&=v_\mathrm{\REV{x}} \sin(\varphi)+v_\mathrm{\REV{y}} \cos(\varphi),\\
    \dot{\varphi}&=\omega,\\
    \label{eqn:final_longitudinal}
    \dot{v}_\mathrm{\REV{x}}&=\frac{1}{m}\left(F_\mathrm{\REV{x}}+F_\mathrm{x}\cos(\delta)-F_\mathrm{f,\REV{y}}\sin(\delta)+m v_\mathrm{\REV{y}} \omega\right),\\
    \label{eqn:final_lat}
    \dot{v}_\mathrm{\REV{y}}&=\frac{1}{m}\left( F_\mathrm{r,\REV{y}}+F_\mathrm{\REV{x}}\sin(\delta)+F_\mathrm{f,\REV{y}}\cos(\delta)-m v_\mathrm{\REV{x}}\omega\right),\\
    \label{eqn:final_omega}
    \dot{\omega}&=\frac{1}{I_\mathrm{z}}\left(F_\mathrm{f,\REV{y}}l_\mathrm{f} \cos(\delta)+F_\mathrm{\REV{x}}l_\mathrm{f} \sin(\delta)-F_\mathrm{r,\REV{y}}l_\mathrm{r}\right),
\end{align}
\end{subequations}

\noindent where\REV{$(p_\mathrm{x},p_\mathrm{y})$} is the position and $\varphi$ is the orientation of the vehicle in the global coordinate frame. The states $v_\mathrm{\REV{x}}$ and $v_\mathrm{\REV{y}}$ denote the longitudinal and lateral velocity of the vehicle in a body-fixed frame and $\omega$ is the yaw rate. The parameters of the model are the distance of the front and rear axis from the center of mass, denoted as $l_\mathrm{f}$ and $l_\mathrm{r}$,  the mass of the vehicle $m$ and the inertia along the vertical axis $I_\mathrm{z}$.

\REV{The longitudinal tire force $F_\mathrm{x}$ is computed using a drivetrain model that assumes a first-order relationship between the motor input and the longitudinal velocity. This modeling approach has been effectively applied to electric vehicles in prior work \cite{Floch22_f1tenth, Liniger_Domahidi_Morari_2015}, therefore, we adopt a similar variant in this study:}
\begin{equation}
    F_\mathrm{\REV{x}}=C_\mathrm{m1}d-C_\mathrm{m2}v_\mathrm{\REV{x}}-C_\mathrm{m3},
\end{equation}
where $C_\mathrm{m1}$, $C_\mathrm{m2}$, $C_\mathrm{m3}$ are lumped drivetrain parameters and $d\in[0,1]$ is the motor input. Lastly, the lateral tire forces are often calculated using a simplified linear tire model as
\begin{subequations}
\label{eq:tire_m}
    \begin{align}
    F_\mathrm{r,\REV{y}} &= C_\mathrm{r}\arctan\left((-v_\mathrm{\REV{y}} + l_\mathrm{r} \omega)/v_\mathrm{\REV{x}}\right),\\
    F_\mathrm{f,\REV{y}}&=C_\mathrm{f}\arctan\left(\delta - (v_\mathrm{\REV{y}} + l_\mathrm{f} \omega)/{v_\mathrm{\REV{x}}}\right),
    %F_\mathrm{r,\eta} &= C_\mathrm{r}\arctan\left(\frac{ -v_\eta + l_\mathrm{r} \omega} {v_\xi}\right),\\
    %F_\mathrm{f,\eta}&=C_\mathrm{f}\arctan\left(\delta - \frac{v_\eta + l_\mathrm{f} \omega}{v_\xi}\right),
    \end{align}
\end{subequations}
where $C_\mathrm{f}$ and $C_\mathrm{r}$ are the cornering stiffness of the front and rear tire, respectively. Finally, the control inputs of the vehicle are the steering angle $\delta$ and the motor input $d$, which a controller can directly actuate.% The physical and empirical parameters of this first principle model can be determined by performing a set of identification experiments \cite{Floch22_f1tenth}.

%\begin{table}[h]
%\caption{Model parameters}
%\label{tab:parameters}
%\begin{center}
%\begin{tabular}{|c|c|c|}
%\hline
%Notation & Value & Unit\\
%\hline
%\hline
%$l_\mathrm{r}$ & 0.168 & m\\
%\hline
%            $l_\mathrm{f}$ & 0.163 & m\\ \hline
%            $m$ & 2.923 & kg \\ \hline
%            $I_z$ & 0.0796 & $\mathrm{kg\cdot m^2}$\\ \hline
%            $C_\mathrm{m1}$ & 41.796& N\\ \hline
%            $C_\mathrm{m2}$ & 2.0152& Ns$\cdot\mathrm{m}^{-1}$\\ \hline
%            $C_\mathrm{m3}$ & 0.4328 & N\\ \hline
%            $C_\mathrm{f}$ & 29.4662 & $\mathrm{N\cdot rad^{-1}}$ \\ \hline
%            $C_\mathrm{r}$ & 41.7372 &$\mathrm{N\cdot rad^{-1}}$\\
%\hline
%\end{tabular}
%\end{center}
%\end{table}

Using the trajectory description of Sec.\ \ref{sec:ref_traj}, we can transform \eqref{eqn:dyn_single_track_model} into a curvilinear coordinate frame (depicted in Fig.~\ref{fig:model} as $\mathcal{F}$) that is parameterized by the position along the reference path \cite{Floch22_f1tenth}:\begin{subequations}
\label{eqn:full_path_following_model}
    \begin{align}
        \dot{s}&=(v_\mathrm{\REV{x}}\cos(\theta_\mathrm{e})-v_\mathrm{\REV{y}}\sin(\theta_\mathrm{e}))/(1-\REV{\kappa}(s)e_s),\\
        \dot{e}_s&=v_\mathrm{\REV{x}}\sin(\theta_\mathrm{e})+v_\mathrm{\REV{y}}\cos(\theta_\mathrm{e}),\\
        \dot{\theta}_\mathrm{e}&=\omega-\REV{\kappa}(s)\dot{s},%\\
    %\dot{v}_\xi&=\frac{1}{m}\left(F_{\xi}+F_{\xi}\cos(\delta)-F_\mathrm{f,\eta}\sin(\delta)+m v_\eta \omega\right),\\
    %\dot{v}_\eta&=\frac{1}{m}\left( F_\mathrm{r,\eta}+F_{\xi}\sin(\delta)+F_\mathrm{f,\eta}\cos(\delta)-m v_\xi\omega\right),\\
    %\dot{\omega}&=\frac{1}{I_z}\left(F_\mathrm{f,\eta}l_\mathrm{f} \cos(\delta)+F_\mathrm{f,\xi}l_\mathrm{f} \sin(\delta)-F_\mathrm{r,\eta}l_\mathrm{r}\right),\\
    %F_\xi&=C_\mathrm{m1}d-C_\mathrm{m2}v_\xi-\mathrm{sign}(v_\xi)C_\mathrm{m3},\\
    %F_\mathrm{r,\eta} &= C_\mathrm{r}\arctan\left(\frac{ -v_\eta + l_\mathrm{r} \omega} {v_\xi}\right),\\
    %F_\mathrm{f,\eta}&=C_\mathrm{f}\arctan\left(\delta - \frac{v_\eta + l_\mathrm{f} \omega}{v_\xi}\right),
    \end{align}
\end{subequations}where the newly introduced states are the position $s$ along the path, the lateral deviation $e_s$ and the heading error $\theta_\mathrm{e}$, while the lateral ($v_\mathrm{\REV{y}}$), longitudinal ($v_\mathrm{\REV{x}}$) and angular velocities ($\omega$) are the same as in \eqref{eqn:dyn_single_track_model}. Furthermore, $\REV{\kappa}(s)$ describes the curvature of the reference path at $s$. 

The main advantage of this model is that the tracking errors explicitly appear in \eqref{eqn:full_path_following_model}, which is beneficial for the control design. Furthermore, due to the physics-inspired model description, all the states can be easily determined from measurements, which allows the design of a full state-feedback controller for the vehicle.

\section{GP-Based Adaptive Control}
\label{sec:adaptive_control}
\vspace{-0mm}\subsection{Control Architecture}
\begin{figure}[t]
      \centering
      \includegraphics[width=1.01\linewidth]{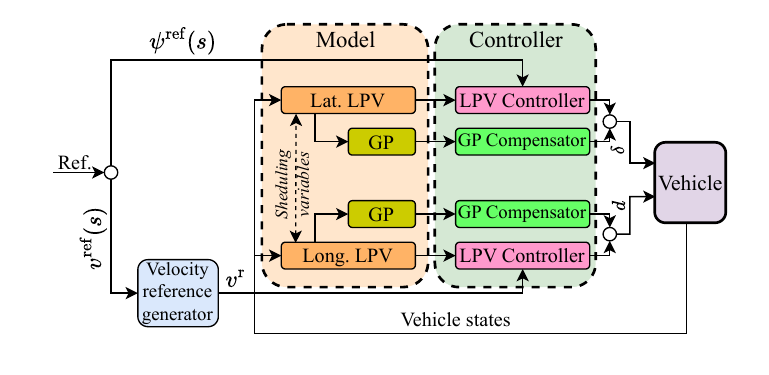}
      %\vspace{-mm}
      \caption{Proposed control architecture for trajectory tracking.}
      \label{fig:control_structure}
      \vspace{-5mm}
   \end{figure}

Based on the trajectory tracking model \eqref{eqn:full_path_following_model}, we propose a computationally efficient adaptive feedback control algorithm. %As \eqref{eqn:full_path_following_model} is a complex nonlinear system for which efficient control laws are hard to derive, 
For this, we decouple the nonlinear vehicle dynamics into two subsystems, corresponding to the longitudinal and lateral motion of the vehicle. Then, to adapt to modeling uncertainties and external disturbances, we augment each subsystem with a GP-based compensator to eliminate the effect of structural model bias. Finally, based on the remaining nominal model, we synthesize LQ state-feedback controllers to track the given reference. The overall architecture is depicted in Fig.~\ref{fig:control_structure}.
\vspace{-2mm}\subsection{Decoupling}
%Following the classical strategy,
We decouple \eqref{eqn:full_path_following_model} into lateral and longitudinal subsystems for individual control design. The longitudinal controller is responsible for tracking the reference velocity and position along the path and the lateral controller is used for path tracking.

From \eqref{eqn:full_path_following_model}, the \textbf{longitudinal model} becomes
\begin{subequations}
     \begin{align}
     \label{eqn:long_s}
        \dot{s}&=(v_\mathrm{\REV{x}}\cos(\theta_\mathrm{e})-v_\mathrm{\REV{y}}\sin(\theta_\mathrm{e}))/({1-\REV{\kappa}(s)e_s}),\\
    \begin{split}
    \label{eqn:long_velocity}
    \dot{v}_\mathrm{\REV{x}}&=\frac{1}{m}\big((1+\cos(\delta))(C_\mathrm{m1}d-C_\mathrm{m2}v_\mathrm{\REV{x}}-C_\mathrm{m3})\\
    &\qquad-F_\mathrm{f,{\REV{y}}}\sin(\delta)+m v_\mathrm{\REV{y}} \omega\big),
    \end{split}
    \end{align}
\end{subequations}
where the state vector $x_\mathrm{lo}=[s\; v_\mathrm{\REV{x}}]^\top$ is the position along the path and the longitudinal velocity, while $d$ is the actuated control input. Furthermore, the heading error $\theta_\mathrm{e}$, lateral deviation $e_s$,  lateral velocity $v_\mathrm{\REV{y}}$,  steering angle $\delta$ and $F_\mathrm{f,\REV{y}}$ are considered as external varying parameters depending on the lateral subsystem. The reference input of the system is $x_\mathrm{lo}^\mathrm{ref}=[s^\mathrm{ref}\; v^\mathrm{ref}]^\top$ defined by the trajectory. To simplify the control design and analysis, we separate the position and velocity states by introducing a virtual velocity generator that provides a position-adjusted virtual reference velocity as $v^\mathrm{r}=v^\mathrm{ref}-k_v(s-s^\mathrm{ref})$. This results in the modified control objective $v_\mathrm{\REV{x}}\rightarrow v^\mathrm{r}$, while the model dimensions are reduced, as only the dynamics of $v_\mathrm{\REV{x}}$ need to be considered. Therefore, the longitudinal behavior can be expressed as:
\begin{align}
\label{eqn:longitudinal_control_model}
\underbrace{\dot{v}_\mathrm{\REV{x}}}_{\dot{\goodchi}_\mathrm{lo}}=&\underbrace{\frac{-C_\mathrm{m2}(1+\cos(\delta))}{m}}_{A_\mathrm{lo}(\delta)} \underbrace{v_\mathrm{\REV{x}}}_{\goodchi_\mathrm{lo}} +\underbrace{\frac{C_\mathrm{m1}(1+\cos(\delta))}{m}}_{B_\mathrm{lo}(\delta)} \underbrace{d}_{u_\mathrm{lo}}\nonumber \\&+\underbrace{\frac{-C_\mathrm{m3}(1+\cos(\delta))}{m}}_{w_0}+\underbrace{\frac{-F_\mathrm{f,\eta}\sin(\delta)}{m}+v_\mathrm{\REV{y}} \omega}_{w_{1}}
\end{align}
where $A_\mathrm{lo}(\delta)$ and $B_\mathrm{lo}(\delta)$ can be considered as parameter-varying state-transition and input matrices with steering angle $\delta$ as a scheduling variable, resulting in a \emph{linear parameter-varying} (LPV) embedding \cite{Toth10_LPV}. In \eqref{eqn:longitudinal_control_model}, $w_0$ is a nonlinearity introduced by the dry friction of the drivetrain and $w_1$ lumps together the effects of the lateral subsystem.

From \eqref{eqn:full_path_following_model}, the \textbf{lateral} behavior of the \REV{car} is described as
\begin{subequations}
\label{eqn:lateral_nonlin_dynamics}
    \begin{align}
\dot{e}_s&=v_\mathrm{\REV{x}}\sin(\theta_\mathrm{e})+v_\mathrm{\REV{y}}\cos(\theta_\mathrm{e}),\\
        \dot{\theta}_\mathrm{e}&=\omega-\REV{\kappa}(s)\dot{s},\\
    \dot{v}_\mathrm{\REV{y}}&=\frac{1}{m}\left( F_\mathrm{r,\REV{y}}+F_\mathrm{\REV{x}}\sin(\delta)+F_\mathrm{f,\REV{y}}\cos(\delta)-m v_\mathrm{\REV{x}}\omega\right),\\
    \dot{\omega}&=\frac{1}{I_\mathrm{z}}\left(F_\mathrm{f,y}l_\mathrm{f} \cos(\delta)+F_\mathrm{\REV{x}}l_\mathrm{f} \sin(\delta)-F_\mathrm{r,\REV{y}}l_\mathrm{r}\right),
    \end{align}
\end{subequations}
where the state vector $x_\mathrm{la}=[e_s\; \theta_\mathrm{e}\; v_\mathrm{\REV{y}}\; \omega]^\top$ consists of the lateral error $e_s$, heading error $\theta_\mathrm{e}$, lateral velocity $v_\mathrm{\REV{y}}$ and yaw rate $\omega$, while $v_\mathrm{\REV{x}}$ is a scheduling variable. The input is steering angle $\delta$. To simplify the model, we first substitute the lateral tire models into \eqref{eqn:lateral_nonlin_dynamics}, use small angle approximations ($\sin(\alpha)\approx\alpha$, $\cos(\alpha)\approx1$), neglect the longitudinal tire force ($F_\mathrm{\REV{x}}\approx0$) and approximate the velocity along the path as $\dot{s}\approx v_\mathrm{\REV{x}}$ which leads to the model \REV{\cite{Rajamani_2012}}:%\cite{Gupta19_lateral_framework}: \vspace{-1mm}
%\begin{equation}
%\label{eqn:lateral_system}
%\begin{split}
%    \begin{bmatrix}
%     \dot{e}_s\\
%    \dot{v}_\eta\\
%        \dot{\theta}_\mathrm{e}\\
%    \dot{\omega}
%    \end{bmatrix}&=\begin{bmatrix}
%        0 & 1 & v_\xi & 0\\
%        0 & -\frac{C_\mathrm{f}+C_\mathrm{r}}{mv_\xi} & 0 & -v_\xi-\frac{l_\mathrm{f}C_\mathrm{f}-l_\mathrm{r}C_\mathrm{r}}{mv_\xi}\\
%        0 & 0 & 0 & 1\\
%        0 & \frac{l_\mathrm{r}C_\mathrm{r}-l_\mathrm{f}C_\mathrm{f}}{I_zv_\xi} & 0 & -\frac{l_\mathrm{f}^2 C_\mathrm{f}+l_\mathrm{r}^2C_\mathrm{r}}{I_zv_\xi}
%    \end{bmatrix}\!\!\!\begin{bmatrix}
%    {e}_s\\
%    v_\eta\\
%    \theta_\mathrm{e}\\
%    \omega
%    \end{bmatrix}\\
%    &+ \begin{bmatrix}
%   0\\
%    \frac{C_\mathrm{f}}{m}\\
%    0\\
%    \frac{l_\mathrm{f}C_\mathrm{f}}{I_z}
%    \end{bmatrix} \delta+\begin{bmatrix}
%        0\\0\\v_\xi\\0
%    \end{bmatrix}c(s),
%    \end{split}
%\end{equation}
\begin{equation}
\label{eqn:lateral_system}
\begin{split}
    \begin{bmatrix}
     \dot{e}_s\\
    \dot{v}_\mathrm{\REV{y}}\\
        \dot{\theta}_\mathrm{e}\\
    \dot{\omega}
    \end{bmatrix}&=\begin{bmatrix}
        0 & 1 & v_\xi & 0\\
        0 & A_{vv} & 0 & A_{v\omega}\\
        0 & 0 & 0 & 1\\
        0 & A_{\omega v} & 0 & A_{\omega \omega}
    \end{bmatrix}\!\!\!\begin{bmatrix}
    {e}_s\\
    v_\mathrm{\REV{y}}\\
    \theta_\mathrm{e}\\
    \omega
    \end{bmatrix}+ \begin{bmatrix}
   0\\
    \frac{C_\mathrm{f}}{m}\\
    0\\
    \frac{l_\mathrm{f}C_\mathrm{f}}{I_z}
    \end{bmatrix} \delta+\begin{bmatrix}
        0\\0\\v_\mathrm{\REV{x}}\\0
    \end{bmatrix}\REV{\kappa}(s),
    \end{split}
\end{equation}
where $A_{vv}=-\frac{C_\mathrm{f}+C_\mathrm{r}}{mv_\mathrm{\REV{x}}}$, $A_{v\omega}=-v_\mathrm{\REV{x}}-\frac{l_\mathrm{f}C_\mathrm{f}-l_\mathrm{r}C_\mathrm{r}}{mv_\mathrm{\REV{x}}}$, $A_{\omega v}=\frac{l_\mathrm{r}C_\mathrm{r}-l_\mathrm{f}C_\mathrm{f}}{I_zv_\mathrm{\REV{x}}}$, $A_{\omega\omega}=-\frac{l_\mathrm{f}^2 C_\mathrm{f}+l_\mathrm{r}^2C_\mathrm{r}}{I_\mathrm{z}v_\mathrm{\REV{x}}}$. Note that in \eqref{eqn:lateral_system} the path curvature is regarded as an external disturbance. Further simplification can be achieved by expressing the lateral model only in terms of the error variables and their derivatives, i.e., by introducing $\ddot{e}_s=\dot{v}_\mathrm{\REV{y}}+v_\mathrm{\REV{x}}\dot{\theta}_\mathrm{e}$.
%Then, as proposed in \cite{Snider09_automatic}, the error dynamics have the form:
%{
%\begin{equation}
%\begin{split}
%    \begin{bmatrix}
%     \dot{e}_s\\
%    \ddot{e}_s\\
 %       \dot{\theta}_\mathrm{e}\\
 %   \ddot{\theta}_e
 %   \end{bmatrix}&=\begin{bmatrix}
 %       0 \!&\! 1 \!&\! 0\! &\! 0\!\\
 %       0 \!& -\frac{C_\mathrm{f}+C_\mathrm{r}}{mv_\xi} \! &\! \frac{C_\mathrm{f}+C_\mathrm{r}}{m} \!&\!-\frac{l_\mathrm{f}C_\mathrm{f}+l_\mathrm{r}C_\mathrm{r}}{mv_\xi}\!\\
 %       0 \!&\! 0 \!&\! 0\! &\! 1\!\\
 %       0 \!&\! \frac{l_\mathrm{r}C_\mathrm{r}-l_\mathrm{f}C_\mathrm{f}}{I_zv_\xi}\! &\! \frac{l_\mathrm{f}C_\mathrm{f}-l_\mathrm{r}C_\mathrm{r}}{I_z}\! &\! -\frac{l_\mathrm{f}^2 C_\mathrm{f}+l_\mathrm{r}^2C_\mathrm{r}}{I_zv_\xi}
 %   \end{bmatrix}\!\!\!\begin{bmatrix}
 %   {e}_s\\
 %   \dot{e}_s\\
 %   \theta_\mathrm{e}\\
 %   \dot{\theta}_\mathrm{e}
 %   \end{bmatrix}\\
 %   &+ \begin{bmatrix}
 %  0\\
 %   \frac{C_\mathrm{f}}{m}\\
 %   0\\
 %   \frac{l_\mathrm{f}C_\mathrm{f}}{I_z}
 %   \end{bmatrix} \delta+\begin{bmatrix}
 %       0\\ \frac{l_\mathrm{r}C_\mathrm{r}-l_\mathrm{f}C_\mathrm{f}}{m} -1\\0\\-\frac{l_\mathrm{f}^2 C_\mathrm{f}+l_\mathrm{r}^2C_\mathrm{r}}{I_zv_\xi}
 %   \end{bmatrix}c(s).
 %   \end{split}
%\end{equation}}
As discussed in \cite{Hu15_heading}, regulating both the heading and the lateral error to the origin yields poor tracking performance, as the two quantities cannot be simultaneously zero along the path if we assume perfect tracking. Therefore, we separate the lateral error and the heading dynamics. We use the lateral error dynamics for feedback control design and only regulate the heading error with a feedforward, similar to the  Stanley controller \cite{Hoffmann07_Stanley}. Furthermore, we also introduce the integral of the lateral error, i.e., $q(t)=\int_0^te_s(\tau)\mathrm{d}\tau$ as a new state to assure asymptotic convergence of $e_s$ without offset. The final control-oriented lateral model can be expressed as

\begin{align}
\label{eqn:lateral_system_2}
    \underbrace{\begin{bmatrix}
        \dot{q}\\
        \dot{e}_s\\
        \ddot{e}_s
\end{bmatrix}}_{\dot{\goodchi}_\mathrm{la}}=&\underbrace{\begin{bmatrix}
        0&1&0\\
        0&0&1\\
        0&0&-\frac{C_\mathrm{f}+C_\mathrm{r}}{mv_\mathrm{\REV{x}}}
    \end{bmatrix}}_{A_\mathrm{la}(v_\mathrm{\REV{x}})}\underbrace{\begin{bmatrix}
        q\\e_s\\ \dot{e}_s
    \end{bmatrix}}_{\goodchi_\mathrm{la}}+\underbrace{\begin{bmatrix}
        0\\0\\ \frac{C_\mathrm{f}}{m}
    \end{bmatrix}}_{B_\mathrm{la}}\delta\\
   +&\underbrace{\begin{bmatrix}
       \!0\!\\
       0\\
       \!\frac{l_\mathrm{r}C_\mathrm{r}-l_\mathrm{f}C_\mathrm{f}}{m} -1\!
   \end{bmatrix}}_{B_\mathrm{\REV{\kappa}}} \underbrace{\REV{\kappa}(s)}_{w_\mathrm{\REV{\kappa}}}+\underbrace{\begin{bmatrix}
        0 \!\!&\!\! 0 \!\!\\
        0 \!\!&\!\! 0 \!\!\\
        \frac{C_\mathrm{r}+C_\mathrm{f}}{m} \!\!&\!\! -\frac{l_\mathrm{f}^2 C_\mathrm{f}+l_\mathrm{r}^2C_\mathrm{r}}{I_\mathrm{z}v_\mathrm{\REV{x}}}
    \end{bmatrix}}_{B_{2}(v_\mathrm{\REV{x}})}\underbrace{\begin{bmatrix}
        \theta_\mathrm{e}\\
        \dot{\theta}_\mathrm{e}
    \end{bmatrix},}_{w_2}\nonumber
\end{align}
where the state vector $\goodchi_\mathrm{la}=[q\;e_s\;\dot{e}_s]^\top$ now only contains the lateral error, its integral and derivative, while $A_\mathrm{la}(v_\mathrm{\REV{x}})$ is the parameter varying state-transition matrix with scheduling signal $v_\xi$ and $B_\mathrm{la}$ is the input matrix. Moreover, $w_\mathrm{\REV{\kappa}}$ is the path curvature and $w_2$ is used to lump together the unmodeled path and heading dynamics.
\vspace{-2mm}\subsection{GP-based Model Augmentation}
\label{sec:GP_augmentation}
Note that we have made simplifications during the derivation of \eqref{eqn:longitudinal_control_model} and \eqref{eqn:lateral_system_2}. Furthermore, due to modeling uncertainties, the model mismatch between the control-oriented model and the true vehicle can significantly decrease the tracking performance. Therefore, to capture this model mismatch, we augment the nominal models \eqref{eqn:longitudinal_control_model} and \eqref{eqn:lateral_system_2} with GPs:
\begin{subequations}
\label{eqn:gp_augmentation}
    \begin{align}
    \label{eqn:gp_augmentation_1}
    \begin{split}
\dot{\goodchi}_\mathrm{la}=&A_\mathrm{la}(v_\mathrm{\REV{x}})\goodchi_\mathrm{la}+B_\mathrm{la}\delta+B_\mathrm{\REV{\kappa}}w_\mathrm{\REV{\kappa}}+B_\mathcal{GP} \mathcal{GP}_\mathrm{la}(\REV{\mathpzc{x}}),\end{split}\\
\begin{split}
\label{eqn:gp_augmentation_2}
\dot{\goodchi}_\mathrm{lo}=&A_\mathrm{lo}(\delta)\goodchi_\mathrm{lo}+B_\mathrm{lo}(\delta)+w_0 + \mathcal{GP}_\mathrm{lo}(\REV{\mathpzc{x}}),
\end{split}
    \end{align}
\end{subequations}
where $\mathcal{GP}_\mathrm{lo}(\REV{\mathpzc{x}})$ and $\mathcal{GP}_\mathrm{la}(\REV{\mathpzc{x}})$ denote the GPs linked to the lateral and longitudinal subsystems, respectively. Based on the structure of the nominal path following model \eqref{eqn:full_path_following_model}, we can observe that the first three equations only capture kinematic relationships. Therefore, we assume that modeling uncertainties only affect the velocity states $v_\mathrm{\REV{x}}$, $v_\mathrm{\REV{y}}$, $\omega$, as proposed in \cite{Hewing18_MPCGP}. Furthermore, we can also note that $B_\mathcal{GP}=[0\;0\;1]^\top$ as uncertain dynamic effects only influence $\dot{e}_s$. 

Observing the original vehicle model \eqref{eqn:dyn_single_track_model}, we can also note that any change in the environmental conditions affects the dynamics through the acting wheel forces. As these models depend on the velocity states, % ($v_\xi$, $v_\eta$, $\omega$), %and the inputs ($d,\delta$)
we choose these variables to construct the GPs. %Therefore, the GP inputs are $z_\mathrm{long}=[v_\xi\;v_\eta\;\omega\;\delta]^\top$ and $z_\mathrm{lat}=[v_\xi\;v_\eta\;\omega\;c]^\top$.
Therefore, the inputs for both the lateral and the longitudinal GPs are $\REV{\mathpzc{x}}=[v_\mathrm{\REV{x}}\;v_\mathrm{\REV{y}}\;\omega]^\top$.

As we previously assumed that all the vehicle states are available, training inputs can be collected from the logged measurement data of driving experiments with the vehicle. By numerical differentiation, we can obtain the state derivatives and the outputs for the GPs can be expressed from \eqref{eqn:gp_augmentation} to generate the training dataset. \TR{Note that process noise effects are handled through the GP estimation process, corresponding to a NARX setting. Colored process noise scenarios can be either handled by appropriate parametrization of the noise covariance/kernel \cite{Rasmussen05_GP} or in case of more elaborate noise settings, using instrumental variables for the mean function estimation \cite{Toth15aAUT}. Furthermore,} because of the large training dataset, utilization of SGPs is necessary to reduce the computation complexity of both the estimation and the online evaluation. 

\vspace{-2mm}
\subsection{Adaptive Control Design}
\label{sec:gp_lpv_lqr}
We propose an adaptive control scheme, which consists of two main components: an adaptive compensator term that accounts for model mismatch, and a nominal offline-designed LPV controller that provides accurate tracking when the model parameters are reliably known.

With the compensator terms, the means of the GPs $\mu_\mathrm{lo}$ and $\mu_\mathrm{la}$ are cancelled by introducing the following compensatory terms:
\begin{subequations}
\label{eqn:GP_feedforward}
    \begin{align}
d_\mathcal{GP}&=1/B_\mathrm{lo}(\delta)\mu_\mathrm{lo}(\REV{\mathpzc{x}}),\\
\delta_\mathcal{GP}&=B_\mathrm{la}^\dagger B_\mathcal{GP}\mu_\mathrm{la}(\REV{\mathpzc{x}}). 
    \end{align}
\end{subequations}
The adaptivity of \eqref{eqn:GP_feedforward} comes from the recursive estimation and compensation with the GPs. Furthermore, the posterior variance characterizing the uncertainty of the GP approximation is utilized to systematically collect informative training data\REV{, which is discussed in Sec. \ref{sec:BO-planning}.}

{Assuming that \eqref{eqn:GP_feedforward} can eliminate the model mismatch, we stabilize the subsystems by the following control laws:\begin{subequations}
\label{eqn:control_laws}
    \begin{align}
    \label{eqn:delta_nom}
            \delta_\mathrm{nom} =&K_\mathrm{la}(v_\mathrm{\REV{x}})\goodchi_\mathrm{la}-\theta_\mathrm{e}-B_\mathrm{la}^\dagger B_\REV{\kappa}w_\REV{\kappa},\\
    \begin{split}
    \label{eqn:d_nom}
            d_\mathrm{nom}=&K_\mathrm{lo}(\delta)(v_\mathrm{\REV{x}}-v^\mathrm{r})+1/B_\mathrm{lo}(\delta)A_\mathrm{lo}(\delta)v^\mathrm{r}\\&-1/B_\mathrm{lo}(\delta)w_0,
    \end{split}
    \end{align}
\end{subequations}
where $K_\mathrm{la}(v_\mathrm{\REV{x}})$ and $K_\mathrm{lo}(\delta)$ are parameter-dependent gain matrices and the additional terms are used to achieve reference tracking and to cancel out known disturbances. The feedback} matrices are obtained using the nominal subsystems \eqref{eqn:longitudinal_control_model} and \eqref{eqn:lateral_system_2} with the LPV-LQR synthesis proposed in \cite{Wu95_LPV}. In the latter sections of this work, we refer to \eqref{eqn:control_laws} as the \emph{nominal} part of the controller and \eqref{eqn:control_laws} and \eqref{eqn:GP_feedforward} combined as the \emph{adaptive} part of the controller.
%\vspace{2mm}

%\textbf{\emph{LPV-LQR synthesis:}} 
Consider an LPV system in the general form as $\dot{\goodchi}=A(\rho)\goodchi+B(\rho)u$. The optimal parameter dependent state feedback matrix $K(\rho)$ that minimizes the quadratic cost $J_\mathrm{LQ}=\int_0^\infty \goodchi^\top\!(t) Q\goodchi(t)+u^\top\! (t) R u(t)\; \mathrm{d}t$ with $Q\in\mathbb{R}^{n_x \times n_x}$ and $R\in\mathbb{R}^{n_u \times n_u}$ can be obtained by solving the %following 
convex optimization problem:
\vspace{-2mm}
\begin{subequations}
\label{eqn:LPV_LQR_synthesis}
\begin{align}
\max_{K,X,Y} &\quad \mathrm{trace}(X),\\
\textrm{s.t.} 
& \quad X\succ 0,\\
& \quad M %\mathrm{LMI}
(X,Y,\mathcal{Q},\mathcal{R},\rho)\succ0\quad \forall\rho\in\mathbb{G},
\end{align}
\end{subequations}
where $X\in\mathbb{R}^{n_x \times n_x}$,  and the \emph{linear matrix inequality} (LMI) constraint $M(X,Y,\mathcal{Q},\mathcal{R},\rho)$ is defined as 
\begin{equation}
\label{eqn:LMI_constraint}
\begin{bmatrix} -\mathrm{He}(A(\rho)X\!+\!B(\rho)Y(\rho))

& (\mathcal{Q}X+\mathcal{R}Y(\rho))^\top\\(\mathcal{Q}X+\mathcal{R}Y(\rho)) & I \end{bmatrix},
\end{equation}
where the gain matrices of the quadratic cost are encoded in $\mathcal{Q}=[Q^\frac{1}{2}\;0 ]^\top$, $\mathcal{R}=[0\; R^\frac{1}{2}]^\top$ , and $\mathrm{He}(X)=X^\top+X$. Furthermore, $Y(\rho)\in\mathbb{R}^{n_u\times n_x}$ is parameterized as follows:
 \begin{equation}
     Y(\rho)=Y_{0}+\rho Y_{1} + \rho^2Y_{2} + \cdots + \rho^nY_{n}
 \end{equation}
and $\mathbb{G}\subset\Gamma$ is a discrete grid of the scheduling region, used to relax the infinite number of LMI constraints. After solving $\eqref{eqn:LPV_LQR_synthesis}$, the parameter-dependent feedback matrix is obtained as $K(\rho)=Y(\rho)X^{-1}$. The parameter-dependent state-feedback gains $K_\mathrm{lo}$ and $K_\mathrm{la}$ for the longitudinal and lateral subsystems have been obtained using the outlined LPV-LQR synthesis with apriori fixed weighting matrices $Q_\mathrm{lo}$, $Q_\mathrm{la}$, $R_\mathrm{lo}$, $R_\mathrm{la}$. %The performance of the controller can be improved by reducing the uncertainty of the GP approximation. This can be achieved by collecting more training data based on the variance of the posterior distribution of the GP. %This concept is implemented in Section~\ref{sec:simulations}, where the covariance of the GP is used to construct specific training trajectories.
%Since LPV-LQR does not require information from the GPs for the synthesis, we can construct a nominal LPV feedback controller with $\mu_\mathrm{long}=\mu_\mathrm{long}=0$, by solving \eqref{eqn:LPV_LQR_synthesis}, with $Q_\mathrm{lat}=\mathrm{diag}(1,80,0)$, $R_\mathrm{lat}=500$ and $Q_\mathrm{long}=1$, $R_\mathrm{long}=100$ weighting matrices for the lateral and longitudinal controller, respectively. Furthermore, $k_v=-0.1$ gain is chosen for the virtual velocity reference generator, to have slower dynamics than the closed loop trajectory tracking system.
\section{Dynamic Active-Learning for Experiment Design}
\label{sec:BO-planning}
\subsection{Problem Statement}
The main objective of the experiment design is to collect data points where the approximation of the GP is less reliable, i.e., where the GP needs the most improvement. This problem has been addressed by active learning methods \cite{Krause08_GP, Gango19_GP_tuning}, which use the variance of the posterior distribution of the GPs to find new training points. However, these methods cannot be applied in dynamic scenarios such as ours as the input of the GPs corresponds to state variables which means that the vehicle has to be navigated to a certain state configuration for the evaluation. \REV{Thus,} systematic trajectory planning is required, taking into account the system dynamics. \REV{For this,} we propose an experiment design that incorporates the system dynamics into the active learning \REV{procedure} and directly synthesizes trajectories to explore regions where the variance of the posterior distribution of the GPs is high. Then, completing the training dataset with new data collected along the resulting trajectory, the GPs can be retrained. In the next sections, we consider our specific scenario, where the inputs of the lateral and longitudinal GPs are the velocity states ($z=[v_\xi\;v_\eta\; \omega]^\top$) of the nonlinear model \eqref{eqn:dyn_single_track_model}. Given baseline GP models, we utilize their posterior distribution to formulate a numerical optimization problem to obtain trajectories for the refinement of the approximation.

%However, as outlined in Sec. \ref{sec:GP_augmentation}, the GP inputs are the velocity states ($z=[v_\xi\;v_\eta\; \omega]^\top$) of the nonlinear model \eqref{eqn:dyn_single_track_model}, from which the trajectory parameters cannot be calculated explicitly.

\vspace{-3mm}\subsection{Trajectory Parameterization}
\begin{figure}[t]
    \centering
    %\begin{subfigure}[b]{0.49\columnwidth}
    %\subfigure[]{
        %\centering
        \includegraphics[width=.45\linewidth]{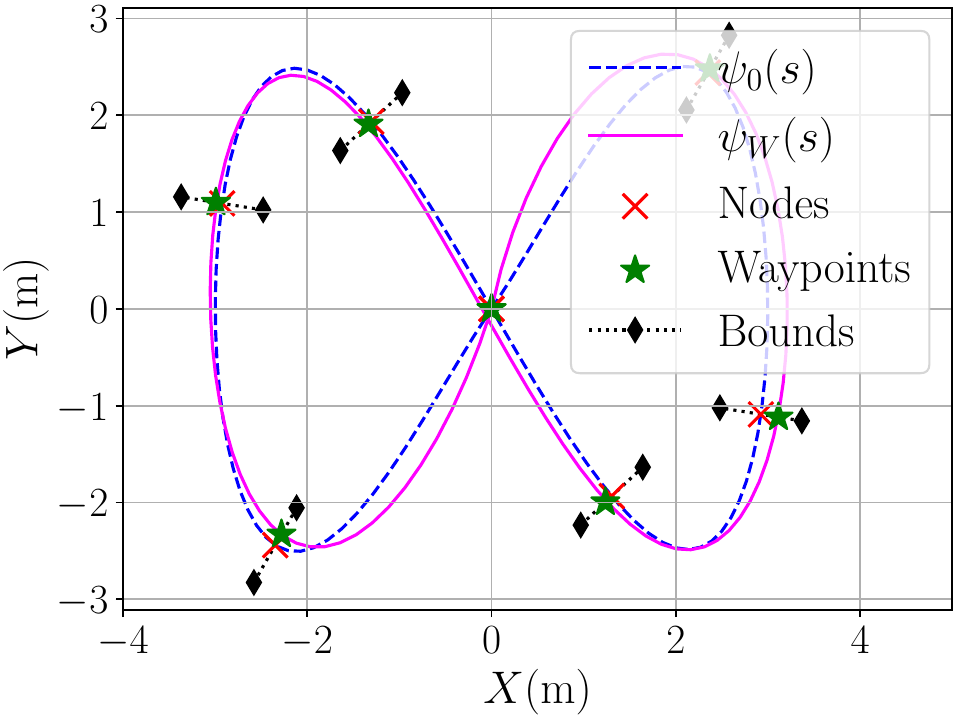}
        %}
    %\end{subfigure}
    \hfill
    %\begin{subfigure}[b]{0.49\columnwidth}
        %\centering
        %\subfigure[]{
        \includegraphics[width=.45\linewidth]{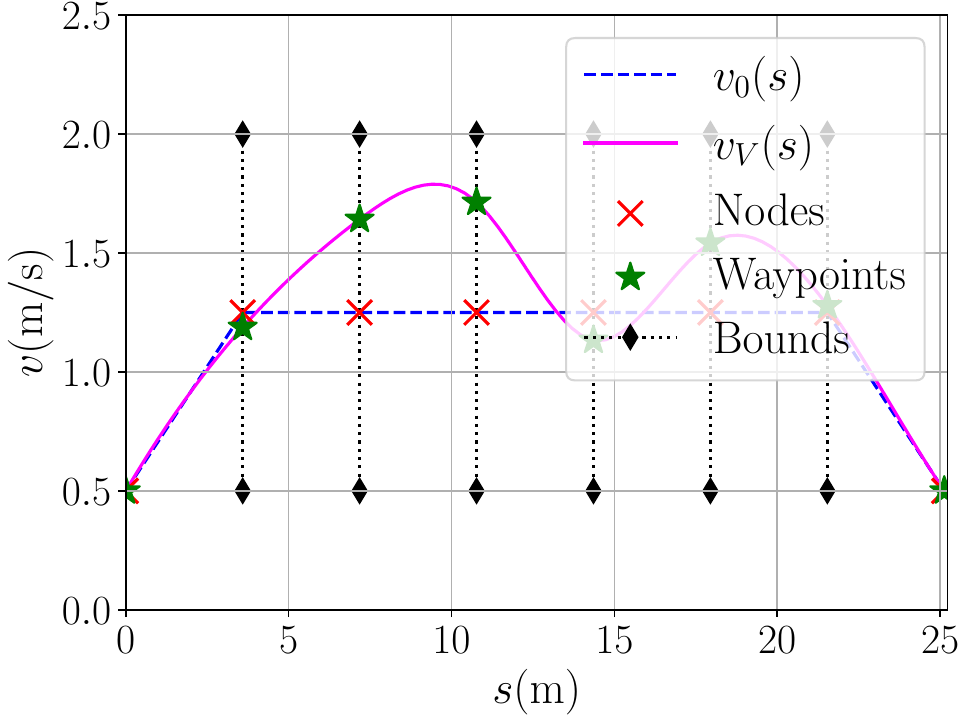}
        %}
    %\end{subfigure}
   \caption[Waypoint-based trajectory parameterization.]{Waypoint-based parameterization of the reference trajectories: the 2D path (left) and the speed profile (right).}
  \label{fig:trajectory_representation}
    \vspace{-2mm}
\end{figure}

The spatial trajectories can be easily parameterized with splines, see Sec.\ \ref{sec:model_and_problem}. However, if all spline parameters are considered free variables, the complexity of solving  a planning task increases significantly, and imposing shape constraints is difficult. Therefore, we adopt a modified version of the parameterization in \cite{Jain20_raceline}. Let an initial trajectory be denoted as $\mathcal{T}_0=\{\psi_0(s), v_0(s)\}$, where $\psi_0(s)$ is a two dimensional arc-length parameterized spline curve and $v_0(s)$ is the corresponding speed profile. For $\psi_0$, we define $n$ number of nodes $\{s_i\}_{i=1}^n$ along this initial path which can be chosen based on various strategies, e.g.\ equidistantly. These are depicted with green dots in Fig.\ \ref{fig:trajectory_representation}. Then, we define new waypoints by moving the node by ${\mathpzc{\REV{w}}}_i$ along a line perpendicular to the $\mathcal{T}_0$ trajectory, where $\REV{\mathpzc{w}}_i\in[-\REV{\mathpzc{w}}_\mathrm{b}/2,\REV{\mathpzc{w}}_\mathrm{b}/2]$ interval. These waypoints are depicted as red crosses. Then, using 2D spline interpolation on the waypoints, we can obtain the path $\psi_W(s)$, which is parameterized by $W=\left[\REV{\mathpzc{w}}_1\dots\REV{\mathpzc{w}}_n\right]^\top$ and the initial trajectory $\psi_0$, as depicted in Fig.\ \ref{fig:trajectory_representation}.

A similar approach can be used for the speed profiles. We consider a generic initial speed profile $v_0(s)$. Then, we define $n$ nodes along $s$ \REV{similarly to} the spatial coordinates, depicted in Fig.\ \ref{fig:trajectory_representation}. By perturbing the velocity profile by $\tilde{v}_i\in[v_\mathrm{min}-v_0(s_i),{v}_\mathrm{max}-v_0(s_i)],\; i\in\mathbb{I}_1^n$\REV{,} at each node, we can define profile points\REV{. Then,} by using spline interpolation, obtain a new speed profile $v_V(s)$ directly parameterized by $V=[v_0(s_1)+\Tilde{v}_1\dots v_0(s_n)+\Tilde{v}_n]^\top$. The main benefit of using this parameterization is that a trajectory can be described by $2n$ number of parameters, which is a significant reduction compared to the full spline parameterization. Furthermore, all parameters have clear physical interpretation, which can be used to guarantee that a trajectory is feasible for the vehicle by altering the bounds $\REV{\mathpzc{w}}_\mathrm{b}$ and $\tilde{v}_\mathrm{b}$.

\subsection{Objective Function Formulation}

To formulate the planning task as a numerical optimization problem, we define the objective function as follows. %First, we create copies of the lateral and longitudinal GPs, $\hat{\mathcal{GP}}_\mathrm{lo}$ and $\hat{\mathcal{GP}}_\mathrm{la}$ respectively. The original $\hat{\mathcal{GP}}_\mathrm{lo}$ and $\hat{\mathcal{GP}}_\mathrm{la}$ are used by the compensator terms \eqref{eqn:GP_feedforward}, while the copies are utilized for the cost function evaluation. Then
First, we drive a simulated vehicle model with the previously trained controller along a trajectory candidate. The simulation is performed in discrete time, i.e. $x_{k+1}=f_\mathrm{sim}(x_k)$, where $f_\mathrm{sim}$ is the nominal nonlinear vehicle dynamics \eqref{eqn:full_path_following_model}, discretized by 4th order Runge-Kutta algorithm.
%can be derived either by discretizing the dynamics in \eqref{eqn:dyn_single_track_model} or by using physics-based simulation models, as described in Sec.\ \ref{sec:simulations}.
During the simulation\REV{,} at each time instant $k$, we observe the GP inputs $\REV{\mathpzc{x}}_k=[0_{3\times3} \; I_{3\times 3}]x_k$ and evaluate the variances of the ${\mathcal{GP}}_\mathrm{lo}$ and ${\mathcal{GP}}_\mathrm{la}$, given by \eqref{eqn:gp_cov}. These variances are denoted as $\hat{\Sigma}_\mathrm{lo}(\REV{\mathpzc{x}}_k, X_{k}),\; \hat{\Sigma}_\mathrm{la}(\REV{\mathpzc{x}}_k, X_{k})$ to highlight that the variance at $k$ only depends on the current input $\REV{\mathpzc{x}}_k$ and the inputs of the training dataset $X_k$. After the evaluation, we accumulate the variances into a single variable, defined as 
\begin{equation}
\label{eqn:J_bayesian_opt}
    J_\mathcal{GP}\REV{(V,W)} = \sum_{k=0}^{N_\mathrm{sim}} (\hat{\Sigma}_\mathrm{lo}(\REV{\mathpzc{x}}_k, X_{k}) + \hat{\Sigma}_\mathrm{la}(\REV{\mathpzc{x}}_k, X_{k})),
\end{equation}
\REV{where the longitudinal and lateral covariances are evaluated by rolling out the controller along a trajectory defined by $V$ and $W$.} Finally, we extend the datasets of the GPs by adding $\REV{\mathpzc{x}}_k$ at each timestep, i.e.\ ${X}_k={X}_{k-1}\cup \REV{\mathpzc{x}}_k$, while ${X}_0$ is the original input dataset. This approach allows us to formulate an objective function that (a) enforces the vehicle to explore the regions where the GP uncertainty is high, (b) accounts for already explored regions, as the addition of new data points reduces the predictive variance in the neighboring regions of $\REV{\mathpzc{x}}_k$. %Furthermore, as we only make alterations to the datasets of the copied $\hat{\mathcal{GP}}_\mathrm{lo}$ and $\hat{\mathcal{GP}}_\mathrm{la}$, the original GPs and therefore the control laws are not altered during the evaluation of the cost function.

Note that the simulation is only used during the design as the substitute for the real dynamics. After a trajectory has been designed, the trajectory will be \REV{executed and} tracked with the real vehicle for the data collection\REV{, which will be used to update the GP}.

\vspace{-3mm}
\subsection{Bayesian Optimization}
Using the outlined trajectory parameterization and the objective function we can formulate the following optimization for the experiment design:
\begin{subequations}
\label{eqn:planning_optimization}
\begin{align}
\label{eqn:baes_NLP_cost}
\max_{V,W} &\quad J_\mathcal{GP}\REV{(V,W)}\\
\textrm{s.t.} & \quad {V}\in[{v}_\mathrm{min},{v}_\mathrm{max}],\\
& \quad {W}\in[-\REV{\mathpzc{w}}_\mathrm{b}/2,\REV{\mathpzc{w}}_\mathrm{b}/2],\\
& \quad x_{k+1}=f_\mathrm{sim}(x_k),
 \end{align}
\end{subequations}
However, simulating the motion of the vehicle along a given reference and evaluating GPs while simultaneously extending their datasets is computationally demanding, therefore, the number of objective function evaluations is limited. Furthermore, as the derivative of the $J_\mathcal{GP}$ cannot be analytically computed and the numerical approximation is difficult, gradient-based optimization methods cannot be used. Therefore, to solve \eqref{eqn:planning_optimization}, we utilize Bayesian optimization \cite{frazier2018tutorial} which is a global optimization technique
that is suitable for solving problems with computationally
expensive cost functions.

After solving \eqref{eqn:planning_optimization}\REV{,} the synthesized reference is applied on the vehicle to obtain the new training dataset. Then, \eqref{eqn:planning_optimization} can be solved again with the expanded dataset, achieving dynamic active learning.

\section{$\mathcal{L}_2$-gain Analysis}
\label{sec:L2}
 \TR{In this section, we will analyze the worst-case performance of our entire tool chain for various parameter realizations of the to-be-controlled vehicle. Specifically, we will characterize the induced $\mathcal{L}_2$ gain of the closed-loop system formed by the nonlinear vehicle model \eqref{eqn:full_path_following_model} for a given parameter realization and the GP-based controller (\eqref{eqn:GP_feedforward} and \eqref{eqn:control_laws}) that was adapted by the same \emph{learning strategy} for each instance of vehicle dynamics. By learning strategy, we mean all the components needed to reproduce the learning process, i.e., the active learning method, the training algorithms, and all initial and tuning parameters. For this purpose,     
 an efficient numerical method is proposed to calculate the closed-loop-induced $\mathcal{L}_2$ gain from reference to tracking error.} This quantitative measure is used to analyze the performance and efficiency of the proposed learning strategy \TR{under variations of the vehicle dynamics}.
 
% on different vehicle models. Applying the learning strategy to different vehicle models and calculating the resulting closed-loop $\mathcal{L}_2$ gains gives quantitative information about the range of vehicles where the learning strategy can be effectively applied. %Implementing a learning strategy on different vehicle models, and calculating the $\mathcal{L}_2$ gain of the resulting closed-loop systems helps exploring the vehicle domain where the learning strategy can be successfully applied. 
Formally, we consider the closed-loop model in the following form:
%To give stability and performance certificates of the proposed control approach, we perform the $\mathcal{L}_2$-gain analysis of the nonlinear model \eqref{eqn:full_path_following_model} with the two independently designed control laws in closed-loop. The nonlinear system has two inputs: the velocity reference $v^\mathrm{r}$ and the curvature $c$ of the path. In the following analysis, we focus on how  $v^\mathrm{r}$ and $c$ affect the tracking performance by analyzing the induced $\mathcal{L}_2$ gain from these inputs to the tracking errors.% As a result for the analysis, the nominal nonlinear model is sufficient.
%For analysis, consider the closed loop in the form of
%we define the nonlinear input-output system as
\begin{equation}
\label{eqn:L2_io_model}
\mathcal{S}\left\{
\begin{array}{rl}
      \dot{{x}}=&\!\!\!\!{f}_\mathrm{cl}({x}, w, {\pt \xi})  \\[1mm]
          z=&\!\!\!\! h({x}, w)
\end{array}\right. 
\end{equation}
where ${f}_\mathrm{cl}: \mathcal{X}\times\mathcal{W} \times\Xi \rightarrow\mathcal{X}$ corresponds to the \TR{true vehicle dynamics} \eqref{eqn:full_path_following_model} \TR{with \eqref{eqn:final_longitudinal}-\eqref{eqn:final_omega} and \eqref{eq:tire_m}} interconnected with the lateral and longitudinal controllers \eqref{eqn:control_laws} and \TR{trained} GP-based compensators \eqref{eqn:GP_feedforward}. The argument $\xi$ describes  the vector of uncertain model parameters, \TR{such as drivetrain parameters, cornering stiffnesses, vehicle mass and inertia parameters} that \TR{describe different realizations of the true vehicle and are assumed to be bounded in a closed set $\Xi \subset \mathbb{R}^{n_\xi}$.}
%can take arbitrary values from a known set $\Xi \subset \mathbb{R}^{n_\xi}$. 
\TR{The nominal model is represented by the parameter vector $\xi_0$ in this set, for which the lateral and longitudinal controllers \eqref{eqn:control_laws} are designed based on \eqref{eqn:LPV_LQR_synthesis} with weighting matrices $Q_\mathrm{lo}$, $Q_\mathrm{la}$, $R_\mathrm{lo}$, $R_\mathrm{la}$. When $\xi=\xi_0$, then\REV{,} from the adaptation point of view\REV{,} there is no \REV{model} mismatch, so there is actually no need for GP augmentation and adaptive controller.} %\TR{[I think we should test the overall algorithm in this case. Why do we throw this out by "Thus \eqref{eqn:L2_io_model} is reduced to the nominal closed-loop system."]}

%\TR{This description of parameter uncertainty of the plant is assumed to be normalized an centered, i.e.,} the nominal model is at $\xi=0$ and $\Xi$ is contains the origin in its interior. The adaptive compensator is assumed to be designed for one uncertainty realization, denoted by $\bar\xi$. If $\bar\xi=0$, then from the adaptation point of view model mismatch is considered, so there is no need for GP augmentation and adaptive controller. Thus \eqref{eqn:L2_io_model} is reduced to the nominal closed-loop system. 

To shift the equilibrium to the origin, the longitudinal velocity $v_\mathrm{\REV{x}}$ and velocity reference $v^\mathrm{{r}}$ are centered: let $\tilde{v}_\mathrm{\REV{x}}=v_\mathrm{\REV{x}}-v^\mathrm{r}_\mathrm{cent}$ and $\tilde{v}^\mathrm{r}=v^\mathrm{r}-v^\mathrm{r}_\mathrm{cent}$, where $v^\mathrm{r}_\mathrm{cent}=(v^\mathrm{r}_\mathrm{max}+v^\mathrm{r}_\mathrm{min})/2$. With these new variables\REV{,}  
${x}=[q\;e_s\;\theta_\mathrm{e}\;\tilde{v}_\mathrm{\REV{x}}\;v_\mathrm{\REV{y}}\;\omega]^\top\in\mathcal{X}$ is the state vector, $w=[ \tilde{v}^\mathrm{r}\;c]^\top\in\mathcal{W}$ is the generalized disturbance signal and $z$ is the generalized performance signal that contains the tracking errors as $z=h(x,w)=[{\pt\tilde{v}}_\mathrm{\REV{x}}-{\pt\tilde{v}}^\mathrm{r}\;\;e_s]^\top\in\mathcal{Z}$. %Note that both the state and the disturbance input are centered such that $\tilde{v}_\xi=v_\xi-v^\mathrm{r}_\mathrm{cent}$ and $\tilde{v}^\mathrm{r}=v^\mathrm{r}-v^\mathrm{r}_\mathrm{cent}$ in order to achieve $w\in\mathcal{L}_2$ and $0={f}_\mathrm{cl}({x}, w)$ equilibrium. Furthermore, to characterize that $v^\mathrm{r}$ is only active in the low-frequency range, we augment the system with a first-order low-pass filter. By determining the $\mathcal{L}_2$-gain of the resulting system from $w$ to $z$, we can provide a quantitative measure for the reference tracking performance of the proposed algorithm. 
Furthermore, to indicate that $\tilde v^\mathrm{r}$ is only active in the low-frequency range, we augment the system with a first-order, strictly proper, low-pass filter. As a result, the state space is extended by one extra dimension, corresponding to the state of the filter i.e.\ $\tilde{x}=[q \; e_s \; \theta_\mathrm{e} \; v_\mathrm{\REV{y}} \; \tilde{v}_\mathrm{\REV{x}} \; \omega\; {\pt \tilde v^\mathrm{r}_\mathrm{f}}]^\top\in\tilde{\mathcal{X}}$. {\pt %The filtered velocity reference is denoted by $\tilde v^r_\mathrm{f}$. 
As the filter is strictly proper, the direct feed-through is eliminated from the output equation}, i.e., $z=h(\tilde{x})=[\tilde{v}_\mathrm{\REV{x}}-{\pt \tilde v^r_\mathrm{f}}\;\;e_s]^\top$.

\TR{\REV{Given a choice of $\xi \in \Xi$, first} an initial sparse GP model is fitted, via the VFE method to obtain the hyperparameters and $M$ inducing points, using a data set with $N$ samples obtained from the nonlinear vehicle model represented by $\xi$ under an input signal $u$ \REV{(fixed for all choices of $\xi \in \Xi$)}. %that is fixed among all trials. 
Then, $N_\mathrm{act}$ iterations of dynamic active learning in terms of experiment design and batch-wise adaptation of the sparse GP are accomplished according to Secs. \ref{sec:GP_update} and \ref{sec:BO-planning} with the hyperparameters fixed among trials. This results in a learned compensator \eqref{eqn:GP_feedforward} that is added to the nominal control law to form the adapted closed-loop system.}

The calculation of the $\mathcal{L}_2$-gain \TR{of the resulting closed loop} relies on the theory of dissipative dynamical systems \cite{vanderSchaft}. %{\pt Let $\xi=\bar\xi$ be fixed at a constant value (i.e. the GP based compensator in $\mathcal S$ is designed for uncertainty realization $\bar\xi$). Then, 
\TR{For \REV{the chosen} $\xi$,} the system \eqref{eqn:L2_io_model} is said to be dissipative w.r.t.\ a quadratic supply $s(w,z)=\gamma^2w^\top w -z^\top z$ if there exists a non-negative storage function $V: \tilde{\mathcal{X}}\rightarrow\mathbb R^+$ such that,  in case $V$ is differentiable, the differential dissipation inequality
\begin{equation}
\label{eqn:dde}
    \dot{V}(\tilde{x})\leq \gamma^2w^\top w -z^\top z
\end{equation}
is satisfied for all $(\tilde{x},z,w)$ trajectories of $\mathcal{S}$. If $w$ is restricted to squared integrable signals, i.e.\, $\mathcal{L}_2$, then the induced $\mathcal{L}_2$-gain is the smallest $\gamma$ for which \eqref{eqn:dde} holds, formally:
    $\sup_{w\in\mathcal{L}_2}\{{\|{z}\|_2}/{\|{w}\|_2}\}\leq\gamma.$
 Furthermore, a finite $\mathcal{L}_2$ gain also proves asymptotic stability of the corresponding system (w.r.t.\ a chosen equilibrium point) as $V$ acts as a Lyapunov function if \eqref{eqn:dde} holds strictly \cite{vanderSchaft}. Following this concept, our goal is to find a storage function $V$ and $\gamma$. However, constructing $V$ and $\gamma$ that give a close upper bound on the true gain is difficult for  a general nonlinear system.  

To overcome this difficulty, we propose an iterative, optimization-based approach, inspired by \cite{Morari21_ROA}, which %The algorithm 
consists of two components: a learner and a verifier.
First, the learner is responsible for finding a storage function candidate $V$ and corresponding $\mathcal{L}_2$-gain $\gamma$ by solving a convex optimization problem. To formulate the learner, we restrict ourselves to function candidates in the form of $V(x)=\tilde{x}^\top P(\tilde{x})\tilde{x}$, where $P(\tilde{x})=P_0+P_1\tilde{x}_1+\cdots+P_7\tilde{x}_7\succ0$ as this function class remains linear in the parameters \REV{$\{P_i\}_{i=0}^7\in\mathbb{R}^{7\times 7}$}, but is more flexible than the generic quadratic forms.

By substituting the storage function and \eqref{eqn:L2_io_model} into \eqref{eqn:dde}:
\begin{multline}
\label{eqn:dde_subbed}
%\begin{split}
    \!\!\!J(\tilde{x},w,{\pt \bar\xi}, \{P_i\}_{i=1}^{7},\gamma^2)\coloneq\tilde{x}^\top P(\tilde{x}) f_\mathrm{cl}{\pt(\cdot)}+f_\mathrm{cl}^\top{\pt(\cdot)} P(\tilde{x}) \tilde{x} \\+ \tilde{x}^\top \frac{\mathrm{d}P(\tilde{x})}{\mathrm{d}t}\tilde{x}- \gamma^2 w^\top w + h^\top(\tilde{x}) h(\tilde{x})\leq0,
\end{multline}
{\pt where $f_\mathrm{cl}(\cdot)=f_\mathrm{cl}(\tilde{x},w,{\pt \bar \xi})$}. Note that if {\pt $\tilde{x}$, $w$} are fixed at constant values, \eqref{eqn:dde_subbed} is linear in the unknown variables $P_i$ and $\gamma^2$. Therefore, by introducing the finite sets $\mathbb{X}\subset\tilde{\mathcal{X}}$ and $\mathbb{W}\subset {\mathcal{W}}$, we propose the following convex optimization problem:
\begin{subequations}
\label{eqn:L2_general_optimization}
\begin{align}
\min_{P_0,\cdots, P_7, \gamma^2} \, & \gamma^2\\
\!\!\!\!\textrm{s.t.}\,\, & P(\tilde{x}) \succ 0,\\
                    \,\, & J(\tilde{x},w, \xi, \{P_i\}_{i=0}^{7},\gamma^2)\leq0,\label{eqn:dde_const}\\ \,\, &\forall (\tilde{x},w)\in\mathbb{X}\times\mathbb{W}, \notag
 \end{align}
\end{subequations}
where $\mathbb{X}$ and $\mathbb{W}$ are discrete grids, constructed by sampling the compact sets such that the sample points sufficiently cover
$\tilde{\mathcal{X}}\times{\mathcal{W}}$. \REV{Note that the denser the generated grid, the more likely it is that the learner can synthesize a valid storage function without any additional sampling.} As \eqref{eqn:dde_const} is linear in the optimization
variables, we can utilize this gridding approach even up to the
case of 6 dimensions, as state-of-the-art numerical solvers
can efficiently handle even a large number of linear constraints. 

Note that the learner only guarantees that the differential dissipation inequality is satisfied at the discrete grid points. Therefore, a verifier is introduced, which essentially tries to find counterexamples where \eqref{eqn:dde_const} does not hold in $\tilde{\mathcal{X}}\times{\mathcal{W}}$ for the previously obtained $\gamma^2$ and $V$. For fixed $P_i$  and $\gamma^2$, the verifier is formulated as the nonlinear optimization:
\begin{equation}
\label{eqn:NLP_optimization}
\max_{\tilde{x}\in\tilde{\mathcal{X}},\;w\in{\mathcal{W}}} \quad J(\tilde{x},w, {\pt\bar\xi}, \{P_i\}_{i=1}^{7},\gamma^2).
 \end{equation}
As \eqref{eqn:NLP_optimization} is a small dimensional problem, numerical solvers with efficient algorithmic differentiation (e.g.\ CasADi \cite{Andersson19_CasADi} - IPOPT \cite{Wachter06_ipopt}) are capable of handling it.  {\pt Note also that since \eqref{eqn:NLP_optimization} is nonlinear, there is no guarantee of finding the global optimum. To avoid getting stuck at a local maximum, \REV{as a workaround, the optimization is performed multiple} times starting from different initial values. For more sophisticated solutions that can provide mathematical guarantees to cover the entire search space, we can combine this simple heuristic approach with scenario method \cite{Campi2018}, or use adaptive sampling similar to that proposed in \cite{Bobiti2018}. \REV{After solving \eqref{eqn:NLP_optimization}, if $J$ is positive, the corresponding $\tilde{x}$ and $w$ values are added to the discrete sets $\mathbb{X}$ and $\mathbb{W}$. The iteration is then repeated until one of two conditions is met: (i) \eqref{eqn:L2_general_optimization} becomes infeasible, indicating that the proposed storage function structure cannot provide a bound for the induced $\mathcal{L}_2$-gain; or (ii) the optimal value of $J$ remains negative, implying that an upper bound for the $\mathcal{L}_2$-gain and storage function $V$ has been found, which also demonstrates the stability of the system.}

\TR{With the proposed approach,} we obtain \TR{the closed-loop} $\mathcal{L}_2$ gain only for \TR{a} specific vehicle model \TR{described} by $\xi$. If we define a sufficiently fine grid $\Xi_\mathrm{g} \subset \Xi$, fix a learning strategy, apply it to all vehicle models $\xi \in \Xi_\mathrm{g}$ and then perform the $\mathcal{L}_2$ gain analysis on each resulting closed-loop system, we can obtain quantitative information on the performance of the controllers achievable under different uncertainty realizations. 

%The whole algorithm is outlined in Alg.~\ref{alg:LP_NLP_iterative_stability}.
%\begin{algorithm}[H]
%\caption{Iterative stability analysis}
%\label{alg:LP_NLP_iterative_stability}
%\begin{algorithmic}[1]
%\State \textbf{input:} $\mathbb{X}$ and $\mathbb{W}$ initial error state-space and disturbance grids
%\Loop
%\State \textbf{solve} \eqref{eqn:incribed_circle_optmization} with $\mathbb{X}$ and $\mathbb{W}$ to obtain $P$
%\If {\eqref{eqn:incribed_circle_optmization} {is unfeasible}} 
%\State \textbf{return} $P=\emptyset$
%\EndIf
%\State \textbf{solve} \eqref{eqn:NLP_optimization} with $P$ to obtain $\max_{\tilde{x},v^\mathrm{r}}(\dot{V}%(\tilde{x})|_{P})$
%\If {$\max(\dot{V}(\tilde{x})|_{P})\geq0$}
%\State \textbf{add} $(\tilde{x},w)=\mathrm{argmax}(\dot{V}%(\tilde{x})|_{P})$ to $\mathbb{X}$ and $\mathbb{W}$ 
%\Else
%\State \textbf{return} $P$
%\EndIf
%\EndLoop
%\end{algorithmic}
%\end{algorithm}

\vspace{-1mm}\section{Simulation Study}
\label{sec:simulations}
\subsection{Setup and Simulation Environment}
\label{sec:satup_and_env}
The performance of the proposed control architecture is first analyzed in a simulation environment. \TR{For the simulation study, we have developed a digital-twin model F1TENTH cars, corresponding to the dynamics discussed in Sec.\ \ref{sec:vehicle_model}, 
using  MuJoCo \cite{Todorov12_mujoco}, a high-fidelity physics engine, which allows us to reliably test the adaptive schemes, as physical parameters can be easily altered. The nominal model corresponds to a parameter configuration of the MuJoCo model, where the parameters were obtained by identifying a real F1TENTH car, see \cite{Floch22_f1tenth}.}
%the nominal model is considered to be dynamic single-track model introduced in Sec.\ \ref{sec:vehicle_model}, with {\pt parameters .} %the parameters taken from \cite{Floch22_f1tenth}. 
The simulation environment and the model parameters are available on GitHub\footnote{\REV{\url{https://github.com/AIMotionLab-SZTAKI/AiMotionlab-Virtual}}}.

To evaluate the performance of the adaptive controller approach, we artificially generate a large model mismatch %by modifying the parameters of the simulated baseline 
{\pt in the digital-twin}. The friction coefficients ($f_\mathrm{wheel}$) between the wheels and the ground are reduced, while their radius ($r_\mathrm{wheel}$) and the overall inertia of the vehicle are increased. The original and altered model parameters are displayed in Tab~\ref{tab:mujoco_params}. Furthermore, the steering dynamics are also altered by scaling and adding an offset to it:
\begin{equation}\label{eqn:steering_gain_and_offset}
\hat{\delta}= c_1\delta+c_0    
\end{equation}
where $\hat{\delta}$ is the steering input acting on the vehicle, $\delta$ is computed by the controller, while $c_1$ and $c_0$ are parameters. %In the case of the nominal model $c_1=1$ and $c_0=0$.%After these changes, there is a significant model mismatch between the nominal model and the simulated car, which justifies the use of the model augmentation.
\begin{table}
    \centering
        \caption{Initial and altered model parameters}
    \begin{tabular}{|c||c|c|c|c|c|}
    \hline
         & $f_\mathrm{wheel}$ & $I_z$ (kgm$^2$) & $r_\mathrm{wheel}$ (m) & $c_1$ & $c_0$ (rad) \\
         \hline\hline
         Initial & 2.5 & 0.078 & 0.052 & 1 & 0 \\
         \hline
         Altered & 0.5 & 0.090 & 0.072 & 0.85 & 0.15 \\
         \hline
    \end{tabular}
    \label{tab:mujoco_params}
\vspace{-2mm}
\end{table}

 For the synthesis of the nominal control laws \eqref{eqn:control_laws} using the baseline single track model, the parameter-dependent gain matrices are obtained by solving \eqref{eqn:LPV_LQR_synthesis}, which has been implemented in Python with CVXPY \cite{Diamond16_CVXPY} and solved with Mosek\footnote{\url{https://www.mosek.com}}. The weighting matrices of the LQR are tuned using numerical simulations with the model resulting in $Q_\mathrm{la}=\mathrm{diag}(1, 80, 0)$, $R_\mathrm{la} = 500$ and $Q_\mathrm{lo} = 1$, $R_\mathrm{lo} = 100$ for the lateral and longitudinal controller, respectively. Furthermore, the gain of the virtual velocity reference generator is $k_v = 0.1$. 

\vspace{-3mm}
\subsection{Data Collection and GP Training}
\label{sec:data_collection_and_training}
For initial GP training, we use lemniscate trajectories (Fig.\ \ref{fig:BO-planner-result}) with constant speeds (0.75-1.25-2 m/s) and 25 Hz sampling, resulting in a dataset of $N={12000}$ points. The initial training of the hyperparameters of the GPs is performed offline by maximizing the VFE cost \eqref{eqn:vfe} by gradient descent with $M=30$. Next, we use the dynamic active learning method from Sec.\ \ref{sec:BO-planning} to refine the GP components. The algorithm uses a lemniscate initial trajectory and a 1.25 m/s constant speed profile is used, with $\REV{\mathpzc{w}}_\mathrm{b}=0.5$, $v_\mathrm{min}=0.5$ m/s, and $v_\mathrm{max}=2$ m/s as bounds. We use $n=7$ equidistant nodes along the path, yielding 14 optimization variables. \REV{These} variables are optimized using 25 initial samples and 20 iterations per trajectory synthesis. The optimization is performed using 25 initial samples and a total of 20 iterations. The trajectories are shown in Fig.\ \ref{fig:BO-planner-result}. 

\begin{figure}
    \centering
    \includegraphics[width=.75\linewidth]{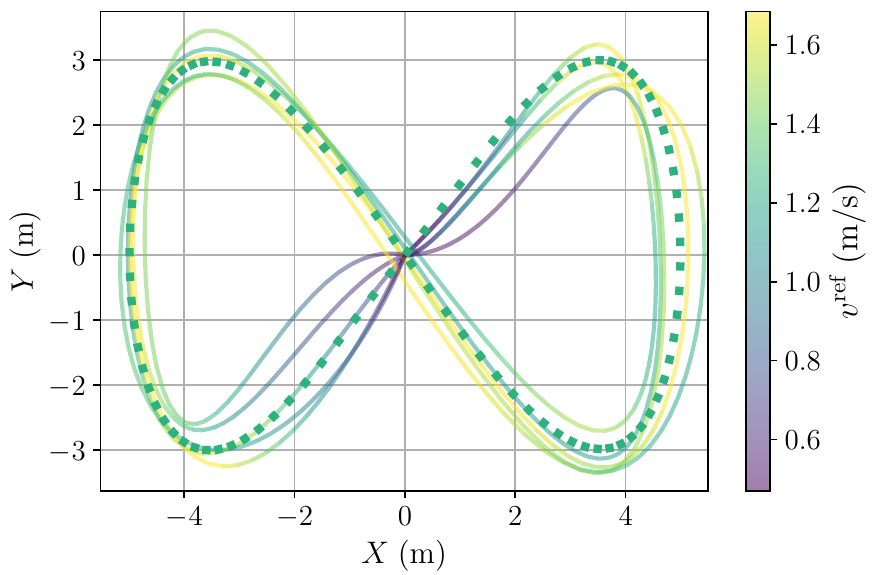}
    \caption[Fine-tuning trajectory.]{Initial training trajectory (dashed) and the trajectories obtained by the dynamic active learning (solid)}
    \label{fig:BO-planner-result}
    \vspace{-1mm}
\end{figure}

Using the data collected along the trajectories, we retrain the GP components in a batch-wise manner with the augmented datasets. We perform 5 iterations of the dynamic active learning, where during each iteration, we collect additional training data along a new trajectory and retrain the GPs. At each iteration, for the sake of the analysis, we evaluate the cumulative variance \eqref{eqn:J_bayesian_opt} along a separate test trajectory to quantify how the active learning method improves the uncertainty of the approximation. As the results show in Fig.\ \ref{fig:variances}, the proposed dynamic active learning efficiently reduces the cumulative variance after each iteration. 

\begin{figure}
    \centering
    \includegraphics[width=.8\linewidth]{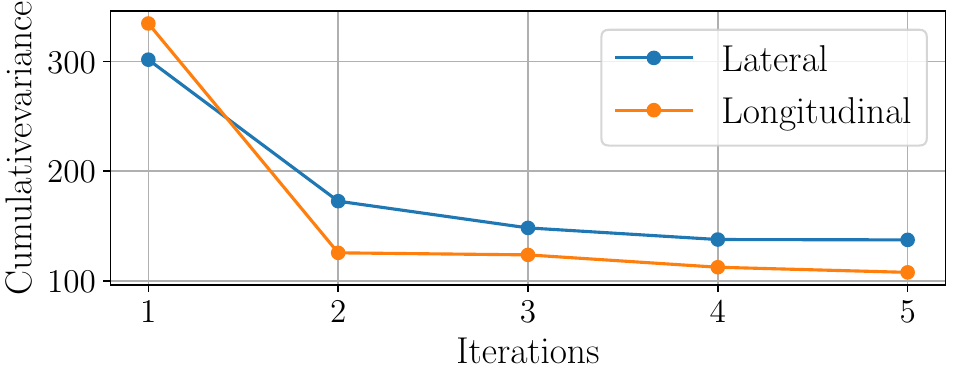}
    \caption{Cumulative variance of the GPs after each iteration of the dynamic active learning algorithm.}
    \label{fig:variances}
    \vspace{-5mm}
    
\end{figure}

\vspace{-2mm}
\subsection{Adaptive Control}
\label{sec:adaptive_control_res}
To demonstrate the efficiency of the proposed control algorithm, we compare \REV{four} scenarios, where the vehicle tracks the same predefined trajectory. In the first scenario (LPV-LQ), only the nominal feedback laws \eqref{eqn:control_laws} are used. Then, in the second and third scenarios, we utilize the GP-based adaptive compensator \eqref{eqn:GP_feedforward} along with the nominal controller. \REV{Finally, in the fourth scenario, we employ the \emph{model predictive contouring control} with a GP-augmented\footnote{\REV{Utilizing the MPCC algorithm without the GP-augmentation resulted in infeasibilities due to the large model mismatch.}} dynamic single track model (MPCC-GP) from \cite{Kabzan19_GPMPC} for comparison. However, it is important to note that the MPCC formulation aims to maximize progress along the path instead of reference trajectory tracking, therefore, for the analysis, longitudinal error measurements are omitted for this scenario.} For online adaptation of the GPs, we start from the offline estimated GP and apply the RLS-GP method of \cite{liu22_learning} and the RGB-GP method of Sec.\ \ref{sec:GP_update} with $Z=20$ batch size. Furthermore, for the RLS-GP $\gamma=0.995$ forgetting factor and $\beta=0.9$ confidence level are used, while for the RGB-GP $n_\alpha=5$ update iterations are selected to maintain computational feasibility, with $\alpha=0.1$ learning rate. \REV{To retain real-time feasibility, we used \emph{real-time iteration} (RTI) for the MPCC scheme.}

The simulation results are \REV{represented in Fig.~\ref{fig:simulations} and Tab.~\ref{tab:fig8_online_results}}. As shown, due to the significant model mismatch, the nominal controller cannot track the reference accurately, as both the lateral ($e_s$), and the heading ($\theta_\mathrm{e}$) errors are significant. Furthermore, a steady-state error occurs in the longitudinal velocity ($v_\mathrm{err}$) that results in increased longitudinal position error ($s_\mathrm{err}$). On the other hand, we can observe that the adaptive controllers can efficiently decrease the tracking errors and ensure the precise tracking of the reference trajectories. A comparison between the control inputs generated by the LPV-LQ and the two GP-based adaptive algorithms is also displayed. Finally, to quantify the tracking performance, the maximum and \emph{root mean square} (RMS) values of the tracking errors are collected in Tab.\ \ref{tab:fig8_online_results}. As the table and the figure show, the RBG-GP provides less noisy estimates and superior tracking performance compared to the RLS-GP. \REV{Furthermore, it can also be seen that the method has similar performance to the MPCC algorithm for a fraction of the computational complexity. This is partially due to the simplicity of the LPV-LQ compared to the MPCC, which requires the solution of an optimization problem at each iteration. Moreover, when dealing with GP-augmented models, the GPs have to be evaluated along the whole control horizon of the MPCC, whereas our proposed solution only requires one evaluation at each control cycle. We furthermore note that, due to the utilization of sparse GPs, the initial offline training data size does not affect the real-time performance of the method.} 

\begin{table}[t]
    \centering
    \caption{Tracking errors \REV{(in meters)} of the nominal, the RLS-based (RLS-GP) and the RGB-based (RGB-GP) online adaptive controllers, \REV{and average control cycle time (in milliseconds).}}
    \begin{tabular}{|c||c|c|c|c|c|}
    \hline
        & max($e_s$) & max$(s_\mathrm{err})$ & $\|e_s\|_\mathrm{RMS}$  & $\|s_\mathrm{err}\|_\mathrm{RMS}$ & \REV{$T_\mathrm{c}$}\\
        \hline
        \hline
        {LPV-LQ}& 0.12 & 1.16 & 0.07 & 0.83  & \REV{0.02}\\ 
        \hline
        {RLS-GP} & 0.05 & 0.37 & 0.01 & 0.15 & \REV{0.77} \\ 
        \hline
        {RBG-GP} & 0.04 & 0.28 & 0.01 & 0.14 & \REV{4.27} \\ 
        \hline
        {GP-MPCC} & 0.06 & -- & 0.02 & -- & \REV{21.14} \\
        \hline
    \end{tabular}
    \label{tab:fig8_online_results}
    \vspace{-3mm}
\end{table}

\begin{figure}
\centering
    \includegraphics[width=\linewidth]{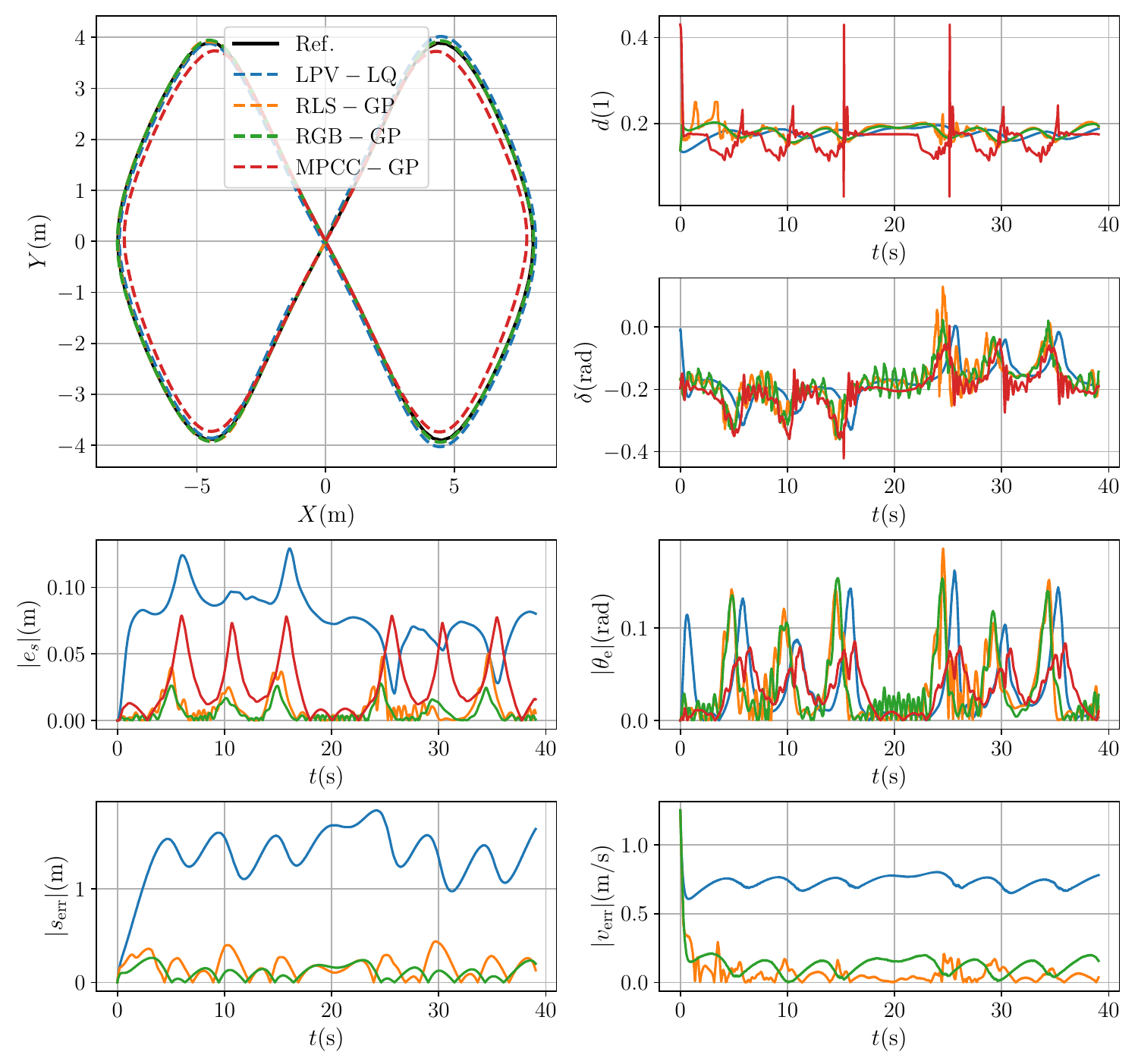}
\caption[Comparison of the nominal and adaptive controllers.]{Comparison of the simulation results using the nominal and the adaptive tracking controllers \REV{as well as the MPCC as a baseline}.}
\label{fig:simulations}
\vspace{-5mm}
\end{figure}

\vspace{-3mm}
\subsection{Performance Verification by $\mathcal{L}_2$ Gain Analysis}
\label{sec:performance_certification}
We \TR{provide performance analysis of} the \emph{learning strategy} of Sec.\ \ref{sec:data_collection_and_training} and the adaptive control scheme of Sec.\ \ref{sec:adaptive_control_res} using the algorithm proposed in Sec. \ref{sec:L2}. For the analysis, we consider the nonlinear closed-loop with the uncertain parameters $\xi=[C_\mathrm{r}\; C_\mathrm{f}\;C_\mathrm{m1}\; C_\mathrm{m2}\; C_\mathrm{m3}\; I_\mathrm{z}]^\top$ as these are the values that are generally harder to identify and can vary based on changing environmental condition. Furthermore, the $\Xi$ set is defined such that each parameter has a maximum 30\% mismatch. We specified the discrete grid $\Xi_\mathrm{g}\in\Xi$ by sampling 4 points equidistantly along each dimension, yielding 4096 realizations. For each realization \REV{of $\xi\in\Xi_\mathrm{g}$}, the $\mathcal{L}_2$-gain computation, \REV{according to the process detailed in Sec.\ \ref{sec:L2},} is implemented in Python, where the learner \eqref{eqn:L2_general_optimization} is formulated in CVXPY and solved with Mosek, while the verifier is handled by CasADi and IPOPT. As discussed in Sec.\ \ref{sec:L2}, a low-pass filter with 10 Hz cutoff is used for shaping the reference which is defined in the range  $v^\mathrm{r}\in[0.5, 2]$ m/s. The region of interest\footnote{$\mathcal{X}\coloneqq[-0.2,0.2]\times[-0.2,0.2]\times[-0.5,0.5]\times[-0.5,0.5]\times[-0.75,0.75]\times[-3.5,3.5]\times[-0.75,0.75]$, $\mathcal{W}\coloneqq[-1.44,1.44, -0.75,0.75]$}
for $\mathcal{X}$ and $\mathcal{W}$ are obtained from simulation results and are discretized with an initial equidistant grid of 5 points along each dimension. In all the cases, the algorithm converged with iteration numbers between {89 and 278}. The resulting upper bounds on the induced $\mathcal{L}_2$ took values in the range $\mathcal{L}_2\in[{0.084}, {0.101}]$. Furthermore, we computed the induced $\mathcal{L}_2$-gain of the nominal closed-loop system, %($\bar{\xi}=0$ with the nominal control laws)
resulting in $\mathcal{L}_2=0.104$. Note that GP augmentation supersedes this, as during the nominal control design, we employed simplifications. From the results, we can conclude that the proposed \emph{learning strategy} can efficiently compensate for the introduced model mismatch under all plant variations, and can restore the original performance.

\TR{Note that the altered digital-twin model we considered in the previous subsections corresponds to the parameter vector $\xi_\mathrm{MJ}=[35.12\;\;23.36\;\;37.98\;\;2.26\;\;0.79\;\;0.09]^\top$, which is in $\Xi$. Hence, our analysis also explains the observed impressive control performance and we can characterize the magnitude of variation of the vehicle dynamics for which we can expect good performance of the learning approach, i.e., when it is safe to deploy the method, which represents the true practical value of the discussed analysis approach.} 

\section{Implementation and Experimental Study}
\label{sec:implementation}
\subsection{Test Environment}
The experiments are performed using the previously introduced F1TENTH \cite{OKelly20_f1tenth} vehicles. For indoor positioning, an OptiTrack motion-capture system is utilized that provides submillimeter position and orientation data. The positioning system and the vehicles are interconnected via Crazyradio PA dongles, which enable low-latency point-to-point communication. The high-level commands and experiment management are achieved through TCP protocol via WiFi. The overall architecture is depicted in Fig.\ \ref{fig:lab-architecture}. The main computation unit of the vehicle is an Nvidia Jetson Orin Nano that runs the control algorithm at 60 Hz, implemented in a ROS2-based onboard software stack available at GitHub\footnote{\url{https://github.com/AIMotionLab-SZTAKI/AIMotionLab-F1TENTH}}.

\begin{figure}
    \centering
    \includegraphics[width=\linewidth]{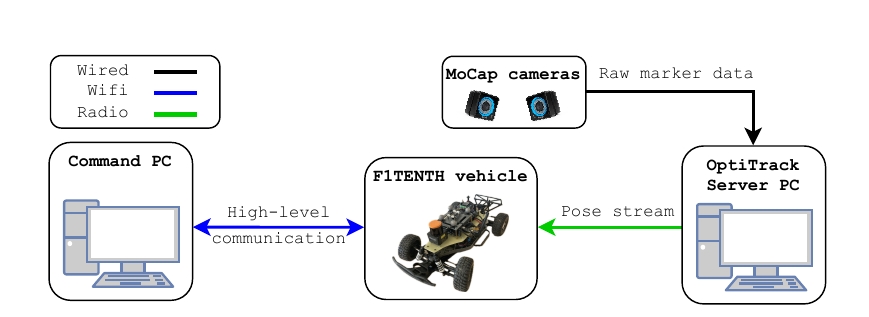}
    \caption{Communication architecture of the test environment.}
    \label{fig:lab-architecture}
\end{figure}

\vspace{-3mm}
 \subsection{Experimental Results}
 The trajectory-tracking performance of the proposed adaptive control method has been evaluated similarly to the simulation studies. First, we generated a significant model mismatch between the previously identified {\pt (nominal)} model %that has been used for the control design 
 and the real vehicle. As shown in Fig.\ \ref{fig:modified_vehicle}, a trailer with $m=2.5$ kg weight is attached to the car. To modify the wheel-ground contact, we used foam instead of the regular carpet of the lab. Furthermore, we modified the steering dynamics by introducing the offset and gain characteristics \eqref{eqn:steering_gain_and_offset}. %Based on our experience, this model mismatch can be significant as the proper identification of the steering dynamics can be time-consuming and requires large space for the experiments \cite{Floch22_f1tenth} that might not be available.

 \begin{figure}
     \centering
     \vspace{-2mm}
     \frame{\includegraphics[width=.45\linewidth]{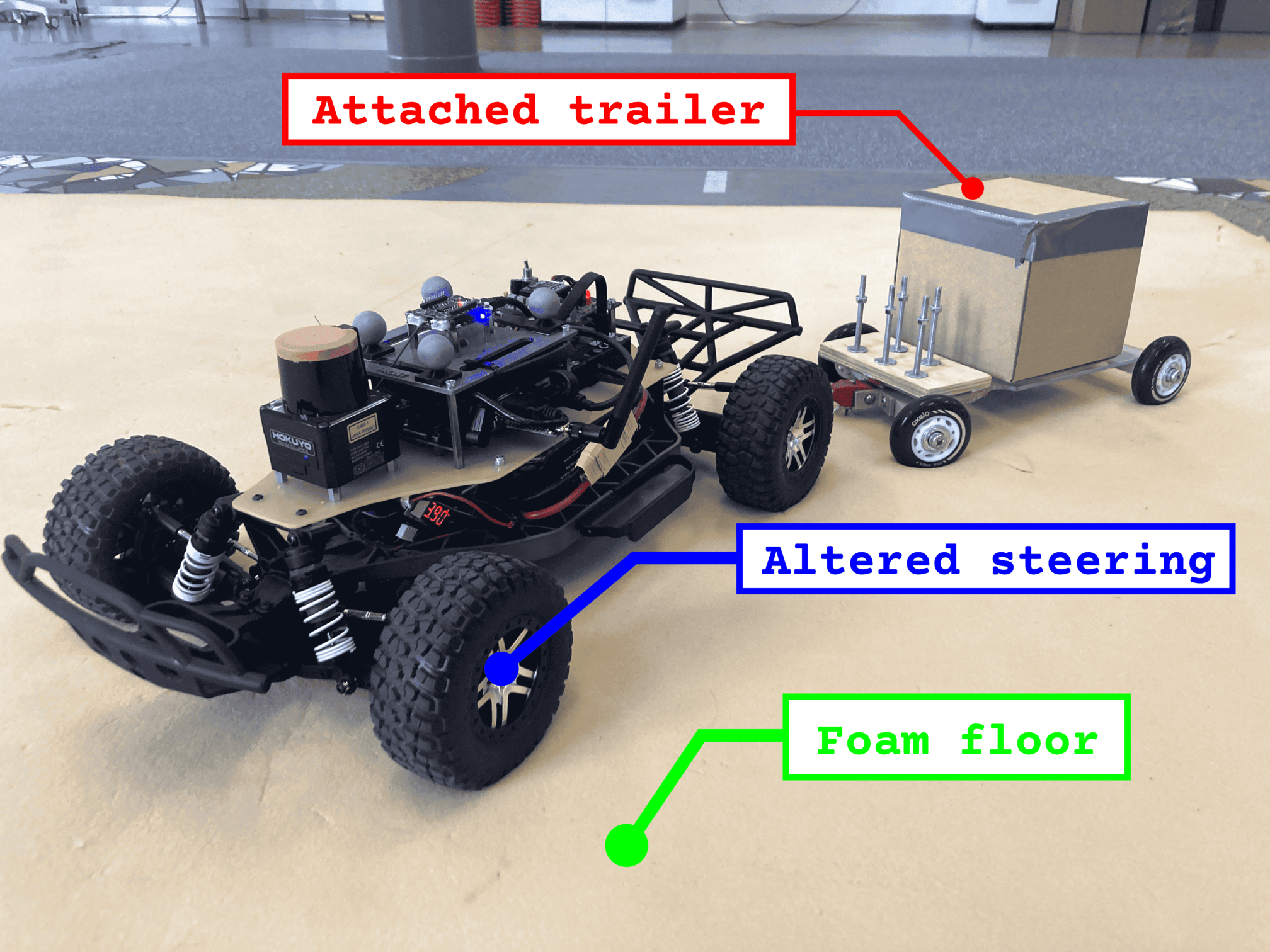}}
     \caption[Modified F1TENTH vehicle platform.]{Modifications made on the F1TENTH platform to artificially generate model mismatch.}
     \label{fig:modified_vehicle}
     \vspace{-5mm}
 \end{figure}

Next, we capture the model mismatch by training the GP components offline. We utilized the lemniscate trajectories of Fig.\ \ref{fig:BO-planner-result} for data collection with constant reference velocities 0.75, 1.25, and 2 m/s as this learning strategy has been validated in simulations. The resulting dataset contained $N=6000$ training points obtained with 25 Hz sampling frequency. To further increase the informative training data, we collected additional samples by executing one iteration of the active-learning-based experiment design. One iteration could sufficiently refine the GPs based on Fig.\ \ref{fig:variances}. For the training, we utilized the SGP regression outlined in Sec. \ref{sec:SGP} with $M=30$ inducing points for both the lateral and the longitudinal subsystems, respectively, similarly to the simulation study. The training of the GP components is carried out with GPyTorch. During the experiments, continuous time state-feedback \eqref{eqn:GP_feedforward}-\eqref{eqn:control_laws} is evaluated at 60 Hz, which is feasible on embedded hardware, as only 30 inducing points are used by the GP components, keeping the control cycle time under 0.01 s.

After the hyperparameters of the GPs had been trained, we analyzed the performance of the controller in an experiment. The measurement results are displayed in Fig.\ \ref{fig:experiment}, where the nominal LPV-LQ (without the GP-based adaptive terms) and the adaptive GP-LPV-LQ are compared. As shown, although, the LPV-LQ has been able to execute the prescribed trajectory, the tracking performance significantly decreased compared to the nominal case and the adaptive controllers, due to the noticeable model mismatch introduced in both the lateral and the longitudinal subsystems.

 \begin{figure}
     \centering
     \includegraphics[width=\linewidth]{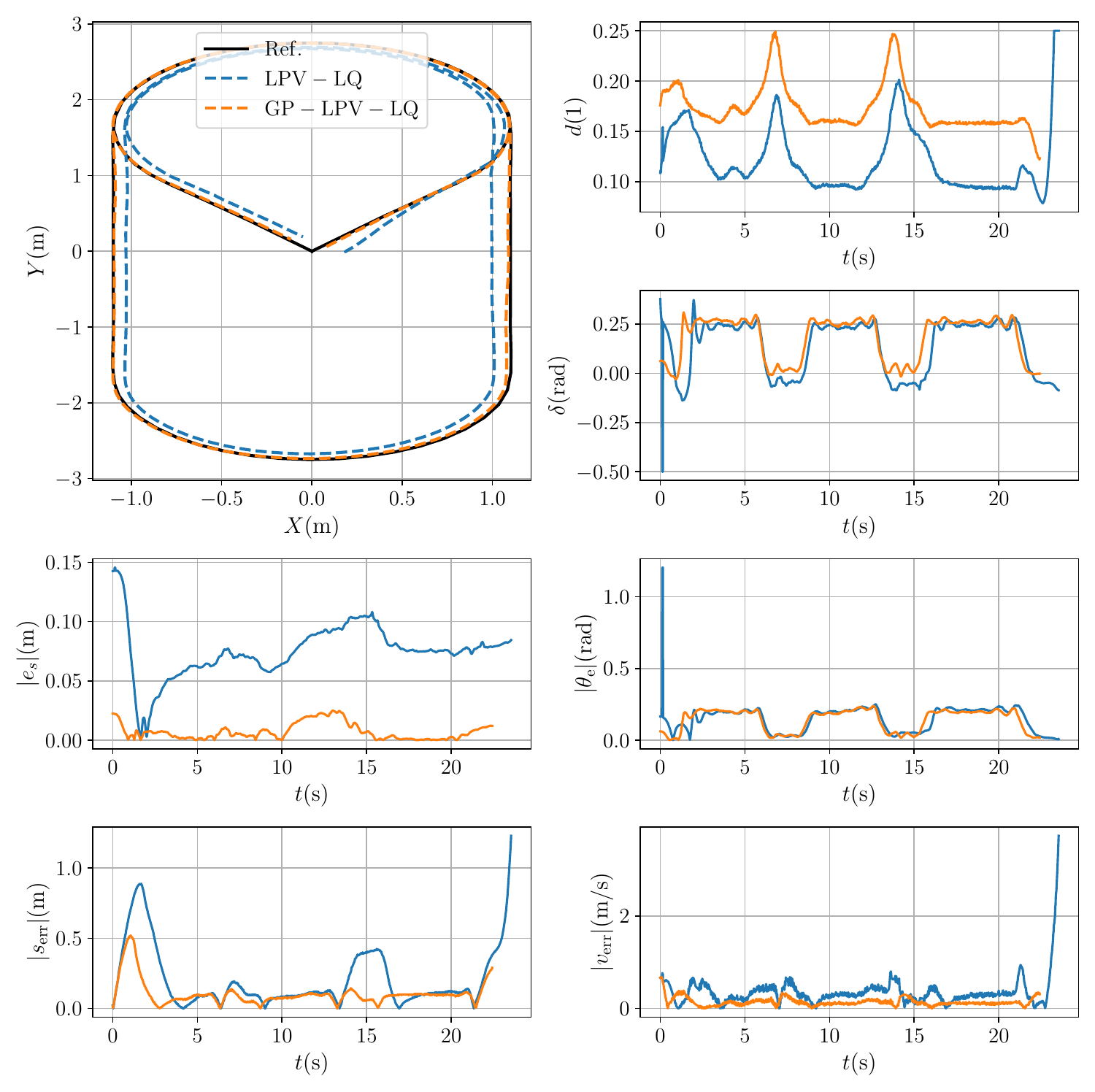}
     \caption{Experimental results of the proposed control scheme.}
     \label{fig:experiment}
     \vspace{-3mm}
 \end{figure}

As the orange lines show, the learning components of the adaptive GP-LPV-LQ have successfully captured the model mismatch and the controller has been able to compensate. A comparison of the maximal and RMS values is presented in Tab.\ \ref{tab:experiment_results}. It is clear that by using the proposed adaptive trajectory-tracking controller, the vehicle has been able to precisely track the reference. However, in both experiments, an initial temporary increase in the longitudinal position error can be observed. Although the GP-based adaptive compensator has reduced the magnitude of the error, it is not negligible. The primary reason for this is that the wheels of the vehicle sink into the soft foam floor, creating an initial sticking effect that the controller needs to overcome. Additionally, a small delay between the prescribed reference and the launch of the control algorithm may contribute to the initial surge in longitudinal error. Nevertheless, it is notable that after 5 seconds, the controller recovered and the performance became satisfactory.

\begin{table}[t]
    \centering
    \caption[Tracking errors in experiment.]{Tracking errors \REV{(in meters)} corresponding to the experiments performed with the F1TENTH vehicle.}
    \begin{tabular}{|c||c|c|c|c|c|c|}
    \hline
        & max$(e_s)$ & max$(s_\mathrm{err})$ & $\|e_s\|_\mathrm{RMS}$ & $\|s_\mathrm{err}\|_\mathrm{RMS}$\\
        \hline
        \hline
        LPV-LQ & 0.11 & 0.89 & 0.08 & 0.29\\ 
        \hline
        \REV{\textbf{GP-LPV-LQ}} & \REV{\textbf{0.02}}  & \REV{\textbf{0.53}} & \REV{\textbf{0.01}}  & \REV{\textbf{0.13}}\\ 
        \hline
    \end{tabular}
    \label{tab:experiment_results}
    \vspace{-5mm}
\end{table}
%\vspace{-3mm}
\section{Conclusion}
{\pt
In this article, a learning-based adaptive control method has
been proposed for autonomous ground vehicles to ensure reliable trajectory tracking in the presence of
modelig uncertainties. By decoupling the nonlinear vehicle dynamics, augmenting the subsystems
with sparse GPs and developing efficient active-learning and recursive sparse GP training methods, we have been able to efficiently capture and compensate for up to 30\% of model mismatch w.r.t.\ a nominal first principle model. The proposed control method has been successfully tested on both a high-fidelity simulator and a real vehicle, demonstrating its practical applicability. To analyze the proposed learning strategy and control architecture,  we have developed a counterexample-based algorithm to estimate the  induced $\mathcal{L}_2$-gain of the
closed-loop system.

%stability and performance properties of the the learning-based control structure we have developed a counterexample-based algorithm to estimate the  induced $\mathcal{L}_2$-gain of the closed-loop system. Although this method is an effective tool to explore the capabilities of a given adaptive controller and training method, it is not applicable to guarantee safe operation during online learning. Developing a safety augmentation around the proposed control structure is one of the major direction for future research.
}
%In this article, an adaptive trajectory tracking approach has been proposed to reliably track reference motion trajectories with autonomous car-like mobile robots. By decoupling the nonlinear vehicle dynamics and augmenting the subsystems with GPs, we have been able to efficiently capture and compensate for up to 30\% of model mismatch. For the data collection for the GP components, we have proposed a dynamic-active-learning-based trajectory planning algorithm, which efficiently reduced the uncertainty of the GPs, leading to more reliable predictions. Furthermore, we have developed a recursive online update method for the GPs to ensure the adaptivity of the controller. Finally to provide performance guarantees for the proposed control scheme, we provided an estimate for the upper bound of the induced $\mathcal{L}_2$-gain of the nonlinear closed-loop system with a counterexample guided iterative algorithm. As demonstrated in the simulation studies and the experimental result, the proposed control approach has been able to provide precise tracking performance by adapting to unmodeled dynamics, while maintaining analytic performance certificates.
\vspace{-3mm}

%\ifCLASSOPTIONcaptionsoff
%  \newpage
%\fi

\bibliographystyle{IEEEtran}
\bibliography{IEEEabrv,references}
\end{document}